\documentclass[twocolumn, 9pt]{aastex631}

\usepackage{mathtools}
\usepackage[mathlines]{lineno} 
\usepackage{amsmath,amsthm,amssymb,bbm}
\usepackage{soul}
\usepackage{comment}
\usepackage{cleveref}
\usepackage{CJK}
\usepackage{appendix}
\usepackage{makecell}
\usepackage{multirow}
\usepackage{subfigure}
\usepackage{tikz}
\usetikzlibrary{bayesnet}
\usetikzlibrary{arrows}
\usetikzlibrary{shapes.misc}
\usetikzlibrary{calc}
\usepackage{pgfplots}
\tikzset{>=latex}
\allowdisplaybreaks
\linenomath
\DeclareRobustCommand{\varlambda}{\text{\usefont{OML}{txmi}{m}{it}\symbol{"15}}}
\def\[#1\]{\begin{equation}\begin{aligned}#1\end{aligned}\end{equation}}
\def\*[#1\]{\begin{align*}#1\end{align*}}
\newcommand{\uat}[2]{\href{http://astrothesaurus.org/uat/#2}{#1 (#2)}}

\begin{document}
\begin{CJK}{UTF8}{gbsn}

\title{Discovery of Two Ultra-Diffuse Galaxies with Unusually Bright Globular Cluster Luminosity Functions via a Mark-Dependently Thinned Point Process (MATHPOP)}

\correspondingauthor{Dayi (David) Li}
\email{dayi.li@mail.utoronto.ca}

\shorttitle{Point Process Model for Inferring Globular Cluster Counts in LSBGs}
\shortauthors{Li et al.}

\author[0000-0002-5478-3966]{Dayi (David) Li (李大一)}
\altaffiliation{Data Sciences Institute Doctoral Fellow,}
\altaffiliation{CANSSI Ontario Multi-disciplinary Doctoral Trainee}
\affiliation{Department of Statistical Sciences, University of Toronto, 700 University Avenue, Toronto, ON M5G 1Z5 Canada}
\affiliation{Data Sciences Institute, University of Toronto, 700 University Avenue, Toronto, ON, M5G 1Z5 Canada}

\author[0000-0003-3734-8177]{Gwendolyn M. Eadie}
\affiliation{Department of Statistical Sciences, University of Toronto, 700 University Avenue, Toronto, ON M5G 1Z5 Canada}
\affiliation{Data Sciences Institute, University of Toronto, 700 University Avenue, Toronto, ON, M5G 1Z5 Canada}
\affiliation{David A. Dunlap Department of Astronomy \& Astrophysics, University of Toronto, 50 St George St, Toronto, ON, M5S 3H4 Canada}

\author[0000-0003-2541-3744]{Patrick E. Brown}
\affiliation{Department of Statistical Sciences, University of Toronto, 700 University Avenue, Toronto, ON M5G 1Z5 Canada}
\affiliation{Center for Global Health Research, St. Michael's Hospital, 30 Bond Street, Toronto, ON M5B 1W8 Canada}

\author[0000-0001-8762-5772]{William E. Harris}
\affiliation{Department of Physics and Astronomy, McMaster University, Hamilton, ON L8S 4M1, Canada}

\author[0000-0002-4542-921X]{Roberto G. Abraham}
\affiliation{David A. Dunlap Department of Astronomy \& Astrophysics, University of Toronto, 50 St George St, Toronto, ON, M5S 3H4 Canada}
\affiliation{Dunlap Institute for Astronomy \& Astrophysics, University of Toronto, 50 St George St, Toronto, ON, M5S 3H4 Canada}

\author[0000-0002-8282-9888]{Pieter van Dokkum}
\affiliation{Department of Astronomy, Yale University, New Haven, CT 06511, USA}

\author[0000-0003-0327-3322]{Steven R. Janssens}
\affiliation{Centre for Astrophysics and Supercomputing, Swinburne University, Hawthorn VIC 3122, Australia}

\author[0000-0001-7549-5560]{Samantha C. Berek}
\altaffiliation{Data Sciences Institute Doctoral Fellow}
\affiliation{Data Sciences Institute, University of Toronto, 700 University Avenue, Toronto, ON, M5G 1Z5 Canada}
\affiliation{David A. Dunlap Department of Astronomy \& Astrophysics, University of Toronto, 50 St George St, Toronto, ON, M5S 3H4 Canada}
\affiliation{Dunlap Institute for Astronomy \& Astrophysics, University of Toronto, 50 St George St, Toronto, ON, M5S 3H4 Canada}

\author[0000-0002-1841-2252]{Shany Danieli}
\altaffiliation{NASA Hubble Fellow}
\affiliation{Department of Astrophysical Sciences, 4 Ivy Lane, Princeton University, Princeton, NJ 08544, USA}

\author[0000-0003-2473-0369]{Aaron J. Romanowsky}
\affiliation{Department of Physics \& Astronomy, San Jos\'{e} State University, One Washington Square,
San Jos´e CA 95192,
USA}
\affiliation{Department of Astronomy and Astrophysics, University of California Santa Cruz, 1156 High Street, Santa Cruz, CA 95064, USA}

\author[0000-0003-2573-9832]{Joshua S. Speagle (沈佳士)}
\affiliation{Department of Statistical Sciences, University of Toronto, 700 University Avenue, Toronto, ON M5G 1Z5 Canada}
\affiliation{Data Sciences Institute, University of Toronto, 700 University Avenue, Toronto, ON, M5G 1Z5 Canada}
\affiliation{David A. Dunlap Department of Astronomy \& Astrophysics, University of Toronto, 50 St George St, Toronto, ON, M5S 3H4 Canada}
\affiliation{Dunlap Institute for Astronomy \& Astrophysics, University of Toronto, 50 St George St, Toronto, ON, M5S 3H4 Canada}



\begin{abstract}
We present \textsc{Mathpop}, a novel method to infer the globular cluster (GC) counts in ultra-diffuse galaxies (UDGs) and low-surface brightness galaxies (LSBGs). Many known UDGs have a surprisingly high ratio of GC number to surface brightness. However, standard methods to infer GC counts in UDGs face various challenges, such as photometric measurement uncertainties, GC membership uncertainties, and assumptions about the GC luminosity functions (GCLFs). \textsc{Mathpop} tackles these challenges using the mark-dependent thinned point process, enabling joint inference of the spatial and magnitude distributions of GCs. In doing so, \textsc{Mathpop} allows us to infer and quantify the uncertainties in both GC counts and GCLFs with minimal assumptions. As a precursor to \textsc{Mathpop}, we also address the data uncertainties coming from the selection process of GC candidates: we obtain probabilistic GC candidates instead of the traditional binary classification based on the color--magnitude diagram. We apply \textsc{Mathpop} to 40 LSBGs in the Perseus cluster using GC catalogs from a \textit{Hubble Space Telescope} imaging program. We then compare our results to those from an independent study using the standard method. We further calibrate and validate our approach through extensive simulations. Our approach reveals two LSBGs having GCLF turnover points much brighter than the canonical value with Bayes' factor being $\sim4.5$ and $\sim2.5$, respectively. An additional crude maximum-likelihood estimation shows that their GCLF TO points are approximately $0.9$~mag and $1.1$~mag brighter than the canonical value, with $p$-value $\sim 10^{-8}$ and $\sim 10^{-5}$, respectively.
\end{abstract}

\keywords{
\uat{Globular Clusters}{656};
\uat{Ultra-Diffuse Galaxies}{940};
\uat{LSBG}{940};
\uat{Point Processes}{1915};
\uat{Astrostatistics}{1882};
\uat{Bayesian Statistics}{1900}
}

\section{Introduction}\label{sec:intro}

Ultra-diffuse galaxies (UDGs) are a class of extended low-surface brightness galaxies (LSBGs) first found in abundance in the Coma cluster by \cite{vanDokkum2015} using the Dragonfly Telephoto Array \citep{Abraham2014}. \cite{vanDokkum2015} defined UDGs through their effective radii ($R_\mathrm{e} > 1.5$~kpc) and $g$-band central surface brightness ($\mu_{0,g} > 24.5$ mag arcsec$^{-2}$).  Subsequently, thousands of UDGs were found in galaxy clusters \citep[e.g.][]{Yagi2016,Wittmann2017,Janssens2019, Lim2020} but also in low-density environments \citep[e.g.,][]{Martinez-Delgado2016,Roman+2019,Forbes+2019,Forbes2020,Danieli+2020}.

Despite their faintness, UDGs seem to have a surprisingly large number of globular clusters (GCs). It is found that UDGs have on average $5-7$ times more GCs than other typical galaxies of the same luminosity \citep{Peng2016, van_Dokkum_2017, Amorisco2018, Lim2018, Danieli2021}. Some even seem to have GC numbers that rival the Milky Way \citep{vanDokkum2016, van_Dokkum_2017}, despite being almost $100$~times fainter. 

It is well-established that there is a monotonic relationship between the stellar mass of the galaxy and the total numbers or mass of the GC population \citep[e.g.,][]{Harris+2013, de2015overlooked,berek2023herbal}, at least for Milky-Way mass galaxies and above. For lower mass galaxies, such as dwarfs, the relationship is somewhat unclear \citep{Eadie_2022, berek2023herbal} since the majority of them have very few or no GCs. Where UDGs fit into this observed relationship is still unclear, and the apparent mismatch between the stellar mass and GC population in UDGs poses an interesting challenge to current models of galaxy formation \citep{Lim2018, VanDokkum2019, Saifollahi2022, Danieli2022, Ferre-Mateu2023, forbes2024ultra, Buzzo2024}. 

UDGs and their GC populations may also provide insight into the nature of dark matter \citep{Hu2000, Hui2017, Wasserman2019, van_Dokkum_2019a}. Multiple theoretical and observational studies have found that there is a relationship between the halo mass of the galaxy and the mass of the GC system \citep{Hudson_2014, Harris_2015, el2019formation, bastian2020globular, chen2023formation}. However, kinematic studies of UDGs have found cases that appear to be extreme outliers from these relations.
For example,
Dragonfly-44 seems to be made almost entirely of dark matter \citep{vanDokkum2016}, while NGC~1052-DF2 and DF4 have little to no dark matter \citep{van2018galaxy, van2019second, Shen2023}. Despite the apparent extremity of the inferred dark matter content of these UDGs, they all have significant GC populations. 

Within the above context, accurate characterizations of GC systems in UDGs are crucial to furthering our understanding of galaxy formation and the nature of dark matter. Thus, inferring the GC counts ($N_{\mathrm{GC}}$) of UDGs has been a central theme in UDG studies \citep[e.g.,][]{Peng2016, Amorisco2018, Lim2018, Lim2020, Carlsten_2022, Saifollahi2022, janssens2024}.

Despite having a larger number of GCs than expected, UDGs/LSBGs still have relatively low $N_{\mathrm{GC}}$ $(\lesssim 100)$ compared to the most massive galaxies in the Universe. Moreover, the majority of UDGs identified to date reside in rather distant galaxy clusters. At these distances, even powerful telescopes such as the \textit{Hubble Space Telescope} struggle to fully probe down to the turnover (TO) point of the globular cluster luminosity function \citep[GCLF;][]{harris+2014}. The combination of small GC counts and limited resolution at large distances makes it difficult to accurately infer and quantify the uncertainty of $N_{\mathrm{GC}}$ in UDGs. The absence of a statistically robust and rigorous GC counting method further exacerbates these challenges.

In many studies \citep[see e.g.,][and others]{Peng2016, van_Dokkum_2017, Lim2018, Saifollahi2022, janssens2024}, different methods and assumptions were used when estimating $N_{\mathrm{GC}}$ of UDGs at large distances, although these methods generally involve the following steps (with some small variations). First, point sources are extracted from astronomical images. Next, GC candidates are identified using the colors and magnitudes of the point sources. Based on the GC candidates, GCs within a certain radius (either predetermined or fitted) from the center of a UDG are then counted. Next, corrections are applied to account for GCs located at larger radii. The count is then adjusted by subtracting an estimated background GC count to correct for chance alignments. Finally, unobserved faint GCs are accounted for by making assumptions about the shape of the GCLF.

Although simple and efficient, the standard method presents various challenges that can impact the accuracy and the uncertainty quantification of $N_{\mathrm{GC}}$ for UDGs:
\begin{itemize}
\item The photometric uncertainties of GC magnitudes and colors increase exponentially as one approaches the detection limit of the images \citep[e.g.,][]{Harris_2023}. Such measurement uncertainties can propagate into the binary selection of GC candidates. Although various studies \citep[e.g.,][]{Peng2016, Harris2020, janssens2024} mitigate this issue by removing sources with high measurement uncertainty, potentially useful information in the data may be discarded, which may impact the analysis.

\item GC membership is uncertain; given a GC candidate, we do not know if it belongs to a UDG, a nearby bright galaxy, or the intergalactic medium (IGM). Intuitively, GCs near the center of a galaxy are more likely to be its member, and the GC spatial distribution in a galaxy is needed to quantify such uncertainty. However, without the GC membership, it is difficult to quantify the GC spatial distribution. How does one quantify such uncertainty heavily affects both $N_{\mathrm{GC}}$ and GCLF estimates. For example, many studies \citep[e.g.,][]{Peng2016, van_Dokkum_2017, Saifollahi2022} estimated the GCLFs of UDGs using only the GC candidates within $1 - 1.5 R_{\mathrm{e}}$ (after a simple constant background subtraction) due to the significant GC membership uncertainty at larger radii. Naturally, this procedure ignores a large amount of data outside the chosen radii and if there are other nearby bright galaxies, a constant background subtraction will bias the results.

\item The chosen radius within which GCs are counted is often fixed and based on the assumption that it contains a certain fraction of $N_{\text{GC}}$ in a UDG \citep[e.g.,][]{van_Dokkum_2017, Lim2018, janssens2024}. This assumption may not always hold, and the associated uncertainties are seldom addressed.

\item To correct for unobserved GCs, various studies adopt a fixed GCLF \citep[e.g.][]{vanDokkum2016, Lim2018, Amorisco2018, Carlsten_2022, Saifollahi2022, janssens2024} where it is either the canonical GCLF or inferred from the stacked GC data within a study. However, the GCLF can vary depending on the host galaxies \citep[e.g.,][]{Villegas_2010}, and the GCLFs of different galaxies may not be the same as that of the ensemble population. Ideally, we would like to infer the GCLF directly from the data for each individual galaxy, but as mentioned, this is challenging if GC membership is unknown, and is further complicated by the photometric measurement uncertainties. Moreover, recent observations \citep[][Tang et al. submitted to ApJ]{Shen_2021, Janssens_2022, Romanowsky_2024} of UDGs NGC1052-DF2, DF4, DGSAT I, and FCC 224 suggest that their GCLFs are weighted towards more luminous GCs than expected from the canonical GCLF. This challenges the widely accepted universality of the GCLF. If true, then this has consequences for any study of UDG GC systems.

\item In many cases \citep[e.g.,][]{Lim2018, Lim2020, Carlsten_2022}, the background subtraction step produces negative $N_{\text{GC}}$ estimates or confidence intervals extending into the negative range, which is unphysical. 
\end{itemize}

Given these complexities, $N_{\mathrm{GC}}$ estimates can vary significantly between different studies of the same UDG. For example, \cite{vanDokkum2016, van_Dokkum_2017, Lim2018, Saifollahi2022, forbes2024ultra} produced $N_{\mathrm{GC}}$ estimates for Dragonfly-44 that range from $20_{-5}^{+6}$ \citep{Saifollahi2022} to $94_{-20}^{+25}$ \citep{vanDokkum2016}. Hence, there is an urgent need for a more statistically robust and rigorous methodology to improve the reliability of $N_{\mathrm{GC}}$ estimates in UDG.

The methods proposed by \cite{Amorisco2018} and \cite{Carlsten_2022} represent significant steps toward accurately estimating $N_{\mathrm{GC}}$ of UDGs. \cite{Amorisco2018} constructed a mixture model to address the GC membership uncertainties while \cite{Carlsten_2022} considered the uncertainty on whether a source is indeed a GC. However, certain limitations persist in their approaches. For example, both \cite{Amorisco2018, Carlsten_2022} assumed a fixed GCLF, which may not account for the variability of GCLFs. \cite{Carlsten_2022} encountered the challenges of negative GC count estimates and confidence intervals. Additionally, neither of these studies accounted for the measurement uncertainties of GC magnitudes and colors. While these studies have offered valuable insights into the challenges of accurately estimating $N_{\mathrm{GC}}$ of UDGs, they also underscore the need for further methodological improvements.

In this paper, we introduce a novel approach to address the challenges of GC counting in UDGs/LSBGs via a hierarchical Bayesian point process model. Point process models have recently garnered attention within the astrophysics community \citep[e.g.,][]{Stein_2015, Li2021, Li2022, Fan_2023, li2024poisson}. As exemplified in \cite{Li2022, li2024poisson}, point processes are natural frameworks to model the data generating process of GCs when trying to discover UDGs through their GC populations. In this work, we further exploit the power of point processes to address the problem of counting the GCs in UDGs.

We propose a Bayesian MArk-dependently THinned POint Process \citep[\textsc{Mathpop};][]{Myllymaki_2009} to jointly model the spatial distribution and the magnitude distribution of GCs. \textsc{Mathpop} is a special point process model where some points are removed (thinned) with probabilities depending on the locations and/or marks (characteristics) of the points. Such a framework seamlessly fits into the context of GC counting: the marks of points are the magnitudes of the GCs, and the unobserved faint GCs are the thinned points in a point process due to their magnitudes and/or locations.

We also take into account the uncertainty when selecting GC candidates. We use two finite mixture models \citep{Benaglia_2009, CHAUVEAU20161} to perform clustering on the color--magnitude data of point sources. Using the clustering results, we construct a probabilistic catalog of GC candidates while incorporating the measurement uncertainties of the point sources. 

The Bayesian approach of our method integrates the probabilistic GC catalog with our \textsc{Mathpop} framework. The joint Bayesian modeling of the GC point pattern and their magnitudes enables us to comprehensively address all associated uncertainties. The point process model gives us the ability to quantify the uncertainty in GC membership assignment, which then provides us the means to model the GCLFs of individual UDGs. In turn, we can also include photometric measurement uncertainty without excluding data. Moreover, through the use of a point process, we avoid unphysical, negative estimates of $N_{\text{GC}}$ and instead obtain the probability that a UDG has no GCs. 

Our paper, which both introduces our method and applies it to 40 LSBGs in the Perseus cluster, is organized as follows. Section \ref{sec:data} introduces the GC catalogs we construct for analysis. Section \ref{sec:method} describes the \textsc{Mathpop} framework and our model. Section \ref{sec:res} contains the results and analysis for the 40 LSBGs in Perseus as well as an extensive simulation study. Section \ref{sec:conclusion} includes discussions and conclusions. A publicly available R package for \textsc{Mathpop} is at \url{https://github.com/davidolohowski/MATHPOP_R_pkg}. Users can also access the \textsc{Mathpop} web-page at \url{https://ddavidli.com/MATHPOP/} for quick \textsc{Mathpop} tutorials as well as code to reproduce the results in this paper.

\section{Data and Background}\label{sec:data}

\subsection{GC Candidates Selection}
We apply our method to data from the Program for Imaging of the PERseus cluster (PIPER; \citealt{Harris2020}) survey. The survey was conducted by the {\it Hubble Space Telescope} ({\it HST}) with its on-board Advanced Camera for Surveys (ACS) and Wide Field Camera 3 (WFC3). The target region for observation is the Perseus galaxy cluster, for which we adopt a distance of $75$ Mpc as in \cite{Harris2020}.

The survey consists of 15 imaging visits, with each visit containing two images within the Perseus region, one captured by ACS and the other by WFC3. These images were taken using the F475W and F814W filters with ACS, and the F475X and F814W filters with WFC3. ACS images span an area of $\sim76\times76$ kpc$^2$ at the Perseus distance, while WFC3 images have an area of $\sim62\times 62$ kpc$^2$. Each image is designated with its visit number and the camera, e.g., V6-ACS is the image obtained by ACS during the Visit 6. 10 of the 15 imaging visits target the outer region of the Perseus cluster at locations of known UDGs, and we use these 10 imaging visits (20 images) to construct the GC catalog.

We note that \citealt{janssens2024} (hereafter J24) conducted an independent study on the GC populations of UDGs via the standard approach using the same PIPER imaging material. We hereafter refer to the GC counting method in J24 as the standard approach. We also apply our method to their GC catalog for comparison, as the GC selection criteria in J24 are different from ours.

In terms of UDGs/LSBGs, the PIPER survey covers 50 LSBGs with 33 of them from \cite{Wittmann2017} (designated W) and 17 from visual inspection of archival CFHT/Megacam imaging by A. Romanowsky (designated R). 23 of the 50 LSBGs meet the strict definition of a UDG \citep{vanDokkum2015} and the rest are either slightly more compact or brighter in surface brightness.

In the next section, we present the procedure to construct our GC catalog, and then briefly discuss the GC catalog from J24.

\begin{figure*}
    \centering
     \includegraphics[width = 0.7\linewidth]{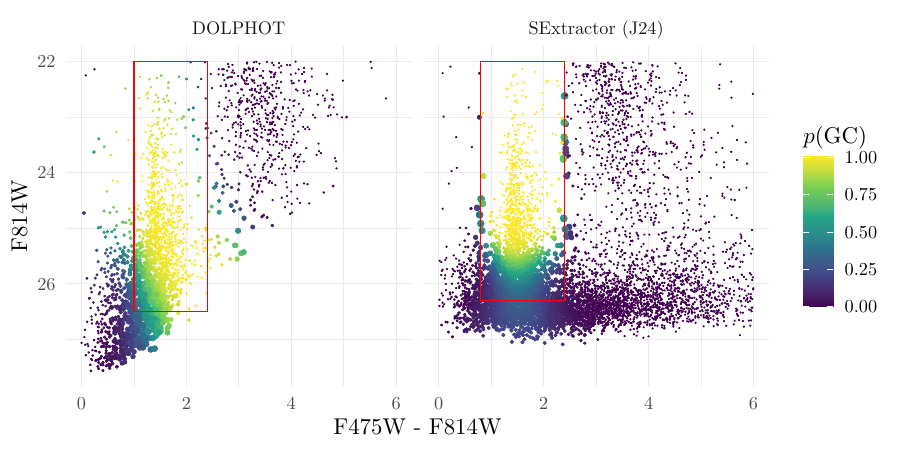}%
     \caption{Color--magnitude diagram of a portion of sources obtained from DOLPHOT and SExtractor (J24). The red boxes indicate the typical selection region of the GCs. The color is the probability that a source is a GC, as obtained from finite-mixture models (see text). The size of the points in both panels indicates the uncertainty of the estimated probability.}
     \label{fig:GC_prob}
\end{figure*}

\subsubsection{Point Source Selection and A Probabilistic GC Catalog}

A major advantage for selecting GCs in Perseus is that, at the 75 Mpc distance, the great majority of GCs appear unresolved (with half-light diameters less than $0.02''$), enabling the use of photometric software codes like DAOPHOT and DOLPHOT. In the original PIPER study \citep{Harris2020}, DAOPHOT \citep{stetson1987} was used for photometry. 

For our study, the 20 images covering the outer Perseus fields were measured anew with DOLPHOT \citep{Dolphin_2000,dolphin2016}.  The procedure to extract the point source list in each field is described in more detail in \cite{Harris_2023}. Briefly, an initial list was constructed with all objects in the field detected in the deepest filter (here F814W), and magnitudes measured by PSF (point spread function) fitting. The list was then culled by including only objects successfully measured in both filters. Next, clearly non-stellar objects of any kind were removed, as determined by the \texttt{sharp} and \texttt{chi} DOLPHOT parameters and magnitude-dependent exclusion boundaries \citep[see particularly the analysis of the central Perseus field for NGC 1275 by][]{Harris_2023}.  A careful manual inspection was performed to confirm and finalize the selection of the point sources. As a final note, the data from images V7-WFC3 and V8-WFC3 were omitted from our analysis, as DOLPHOT could not converge to solutions as it failed to coordinate-match the list of point sources found in these two fields. Due to the convergence issue, 40 of the 50 LSBGs in the PIPER survey were covered in our DOLPHOT data, with one LSB galaxy, W7, being observed in V12-ACS and partially observed in V14-ACS. The structural properties of the LSBGs were obtained by J24, and we use their results for subsequent analyses. Our GC candidate catalog is thus constructed based on a total of 18 images at different locations in the Perseus region.


The final photometric lists are in the natural \textit{HST} filter magnitudes in Vegamag.  
For the WFC3 data, the measurements in $\mathrm{F475X}$ were transformed to the equivalent $\mathrm{F475W}$ (ACS) \citep[see][]{Harris2020}. Foreground extinctions were removed for all measurements and individual fields, as given in \cite{Harris2020}.

The left panel of Figure \ref{fig:GC_prob} shows the color--magnitude diagram (CMD) for a portion of extracted point sources ($50\%$ of data are randomly selected due to memory constraints). 
Because normal, old GCs occupy a well-determined range of magnitudes and colors, corresponding to their luminosities and metallicities, a typical GC selection method is to construct a box that 
generously encloses the expected color and magnitude range for GCs \citep[see][for details]{Harris2020}. This approach is represented by the red box in the left panel of Figure \ref{fig:GC_prob}, which contains likely GCs. This approach is a binary selection, in that point sources are identified as likely GCs or not. 

Rather than applying a strict binary classification to identify GCs, we adopt a probabilistic approach. The motivation is obvious: GCs certainly do not reside  in a box in the CMD. Not only will some GCs reside outside the box, but the box will also contain some number of contaminants (foreground stars or faint, small background galaxies). Moreover, measurement uncertainties in the color--magnitude data can be significant at fainter levels, which can have a strong impact on the binary selection results. Therefore, we apply a non-parametric multivariate finite mixture model \citep{Benaglia_2009, CHAUVEAU20161} to the color--magnitude data to estimate the probability that a source is a GC. The details of obtaining the probability are given in Appendix \ref{Appx A subsec: our data}. 

The color of each point in the left panel of Figure \ref{fig:GC_prob} shows the estimated probability that a source is a GC, while the size represents the uncertainty of the estimates. For example, a point that is yellow and small has a high probability of being a GC, with low uncertainty. Note that we have excluded sources with $\mathrm{F814W}< 22.0$~mag and $\mathrm{F475W} - \mathrm{F814W} < 0.0$~mag to stabilize the clustering algorithm. These sources are also highly unlikely to be GCs.

\subsubsection{Comparison to GC Catalog from J24}

The selection process for the GC catalog by J24 was rather similar to ours (see J24 for more details). Briefly, \textsc{SExtractor} \citep{Bertin1996} was run in dual-image mode on all images using the F814W images, after removing the light profile of LSBGs using the best-fit \texttt{Imfit} \citep{erwin2015imfit} model. A point source catalog was then constructed using the concentration parameter $C_{5-12}$, the F814W magnitude difference measured in 5 and 12 pixel diameter apertures\footnote{Note that J24 used a different set of science images with a final pixel scale of $0.03\arcsec$.}. Lastly, GC candidates were selected from point sources using a binary selection criterion in (extinction corrected) magnitude and color of $21.5 < \mathrm{F814W}< 26.3~\mathrm{mag}$, and $0.8 < \mathrm{F475W}-\mathrm{F814W} < 2.4~\mathrm{mag}$. The faint limit of the selection criterion corresponds to the canonical GCLF TO point. The right panel of Figure \ref{fig:GC_prob} shows the CMD of point sources from J24 (only showing $50\%$ of sources), while the red box in the figure shows the GC selection criteria by J24. 

We also construct a probabilistic GC catalog using the J24 point source list. However, since the point source catalog in J24 contains significantly more contaminants than ours at fainter levels (see Appendix \ref{Appx A subsec: J24 data} for a detailed account of the existence of these contaminants), the method used for our data has issues in separating GCs from contaminants using the color--magnitude data. Thus, we use a different method to obtain the probabilistic catalog from the one used for our data. In short, we consider a two-component parametric mixture model to cluster the color--magnitude data of sources that pass the color cuts by J24 (see Appendix \ref{Appx A subsec: J24 data}). As in the left panel of Figure~\ref{fig:GC_prob}, the color and size of points in the right panel represent the estimated probability and uncertainty.

The difference between the probabilities shown in Figure \ref{fig:GC_prob} for the two datasets is mainly due to identification of point sources in each approach. 
\textsc{SExtractor} is more permissive than DOLPHOT for what is considered a point source in that \textsc{SExtractor} only considered point sources successfully detected in $\mathrm{F814W}$ filter, while DOLPHOT considered those detected in both filters. Thus, at faint levels, the J24 catalog contains more contaminants than our catalog, which explains the significant difference in GC probabilities between the two catalogs around $(1.5\text{ mag}, 26\text{ mag})$ in the CMDs. With only color--magnitude data, no clustering algorithm can confidently separate GCs from contaminants in this region, owing to the large numbers of contaminants (see Appendix \ref{Appx A subsec: J24 data} for further details).

\subsection{GC Removal due to Completeness Fraction}

The main challenge in estimating GC counts in UDGs is that faint GCs are unobservable because of the detection limits in the images. The proportion of detectable GCs is usually a function of their magnitudes $m$, quantified by the completeness fraction, $f(m)$. More generally, the completeness fraction is a function of both the magnitude $m$ and the local background surface brightness, which is a function of location $s$, since brighter background light around the centers of host galaxies makes it more difficult to detect faint objects \citep[see][for examples]{Harris_2023}.  However, in our case we assume $f(m,s) = f(m)$, since the background stellar light from UDGs/LSBGs is extremely faint and the images avoid the brightest core region of the Perseus cluster.

To determine $f(m)$, artificial star tests (AST) \citep[cf.][]{Harris2020,Harris_2023} were performed and the point source recovery rate was measured in the range of $20.0 - 31.75$~mag for F814W and $23-31.5$~mag for F475W. $f(m)$ is obtained by fitting the AST results with a logistic function \citep[see e.g.,][]{Harris+2016, Harris2020}:
$$
f(m) = 1/(1 + \exp(a(m - m_0)),
$$
where $a$ controls the rate of decrease in $f(m)$ as $m$ increases and $m_0$ is the $50\%$ completeness limit. The fitted parameters $a$ and $m_0$ for the two filters F475W and F814W from the AST are included in Appendix \ref{appdx: AST}. 

\subsection{Measurement Uncertainty in Magnitude}

As the brightness of a GC approaches the image detection limit, the uncertainty $\sigma_M$ in its measured magnitude increases. Typically, $\sigma_M(m_t)$ is a function of the true GC magnitude $m_t$, and characterized by an exponential function \citep{Harris_2023}:
\[
\sigma_M(m_t) = \beta_0\exp(\beta_1(m_t - m_1)),
\]
where $\beta_0, \beta_1, m_1$ are free parameters. These parameters are again determined through AST, where measurement uncertainties are derived by comparing the true magnitudes of injected artificial stars with their measured magnitudes from DOLPHOT. During model fitting, we set $m_1=25.5$~mag (roughly the central value of the magnitude range being considered in the AST), and the fitted values from the AST for $\beta_0, \beta_1$ are given in Appendix \ref{appdx: AST}.

\begin{deluxetable*}{llcl}[t]
\centering
\tablehead{\colhead{} & \colhead{Type} & \colhead{Unit} & \colhead{Meaning}}
  \startdata
  $\mathbf{x}$ & Data & - & Observed GC point pattern data \\
  $m$ & Data & mag & Observed GC magnitude data in F814W\\ 
  $\mathcal{S}$ & Constant & - & Observation window (image field of view)\\
  $\mathbf{s}$ & Random variable & (kpc, kpc) & Unrealized random location of GC \\ 
  $s$ & Data/Place-holder & (kpc, kpc) & Location of an observed GC/a fixed location in $\mathcal{S}$ \\
  $\pi$ & Function & - & Probability/probability density function (p.d.f.) \\
  $\mathbf{X}, \mathbf{X}'$ & Random processes & - & Thinned ($\mathbf{X}$) and unthinned ($\mathbf{X}'$) GC point processes \\
  $\Lambda, \Lambda'$ & Latent functions & kpc$^{-2}$ & Intensity of thinned ($\Lambda$) and unthinned ($\Lambda'$) GC point processes \\
  $\lambda_0$ & Model Parameter & kpc$^{-2}$ & Intensity of unthinned GC point process in IGM. \\
  $N_{\text{GC}}$ & Latent variable & - & Number of GCs in a galaxy \\
  $\varlambda$ & Model Parameter & - & Rate parameter for the mean number of GCs in a galaxy \\
  $c$ & Constant & (kpc, kpc) & Central location of a galaxy \\
  $R_h$ & Model Parameter & kpc & Half-number radius of a GC system \\
  $R_{\mathrm e}$ & Constant & kpc & Effective radius of a galaxy \\
  $\alpha$ & Model Parameter & - & S\'{e}rsic index of a GC system \\
  $\vartheta$ & Constant & radian & Orientation angle of a GC system \\
  $e$ & Constant & - & Ellipticity (aspect ratio) of a GC system \\
  $M$ & Random variable & mag & Measured magnitudes in F814W of GCs \\
  $M_{\text{true}}$ & Latent variable & mag & True magnitudes in F814W of GCs \\
  $f(m)$ & Fixed function & - & Completeness fraction \\
  $\sigma_M(m)$ & Fixed function & mag & Measurement uncertainty of GC magnitudes \\
  $\rho(s)$ & Function & - & Spatial thinning probability \\
  $\mu_{\text{TO}}$ & Model Parameter & mag & GCLF TO point in F814W (not adjusted for distance) \\
  $\sigma$ & Model Parameter & mag & GCLF dispersion \\
  $\phi(m; \mu_{\text{TO}}, \sigma^2)$ & Latent Function & - & True GCLF \\
  $\psi(m; \mu_{\text{TO}}, \sigma^2)$ & Latent Function & - & GCLF with measurement uncertainty \\
  $\mathcal{N}_f(\mu_{\text{TO}}, \sigma^2)$ & Distribution & mag & Distribution of observed GC magnitudes in F814W \\
  $\Psi_f$ & Latent variable & - & Proportion of GCs that are observable given $f(m)$ and GCLF \\
  \enddata
\caption{Parameters and notations used in this paper. Subscripts and superscripts are dropped for simplicity. Note that model parameters are quantities we directly infer about the model. Random variables and processes are quantities directly related to data generation. Latent variables and functions are intermediate unknown quantities determined by other quantities such as model parameters and random variables.} 
\label{tab:parameter table}
\end{deluxetable*}

\section{Methods}\label{sec:method}

In this section, we introduce our methodology, \textsc{Mathpop}, for inferring GC counts using point processes. While traditional methods to infer GC counts require fragmented steps and various assumptions, modeling GCs through point processes is a coherent approach that keeps assumptions to a minimum. Readers familiar with point process models may skip to Section \ref{sec:MTPP}, where we introduce the concept of a mark-dependently thinned point process. For quick reference, Table \ref{tab:parameter table} contains the list of model parameters and mathematical notation. 

\subsection{Spatial Point Process}\label{subsec:PP}

The simplest spatial point process is a homogeneous Poisson process (HPP) with a constant intensity $\lambda \geq 0$, independent of locations $s$. If the intensity $\lambda(s)$ changes with $s$, then we instead have an inhomogeneous Poisson process (IPP). In our application, the intensity is the mean point count per unit area, and we model the GCs as points generated by an IPP, as follows.

We begin by considering the simplest scenario where we have observed locations $\mathbf{x} = \{s_i\}_{i=1}^n$ of GCs in one UDG. We assume that $\mathbf{x}$ is generated by a GC point process $\mathbf{X} = \{\mathbf{s}_i\}_{i=1}^n$, where $\mathbf{X}$ resides in a bounded observation window $\mathcal{S} \subseteq \mathbb{R}^2$, which is the field of view of an image. The main characteristic of $\mathbf{X}$ is its intensity function $\lambda(s) \geq 0, s \in \mathcal{S}$.

Importantly, we purposely define $\mathbf{X}$ for the observed GCs only. In this way, we can account for incompleteness in GC counts through a thinned point process, described next.


\begin{figure*}
    \centering
    \includegraphics[width = \textwidth]{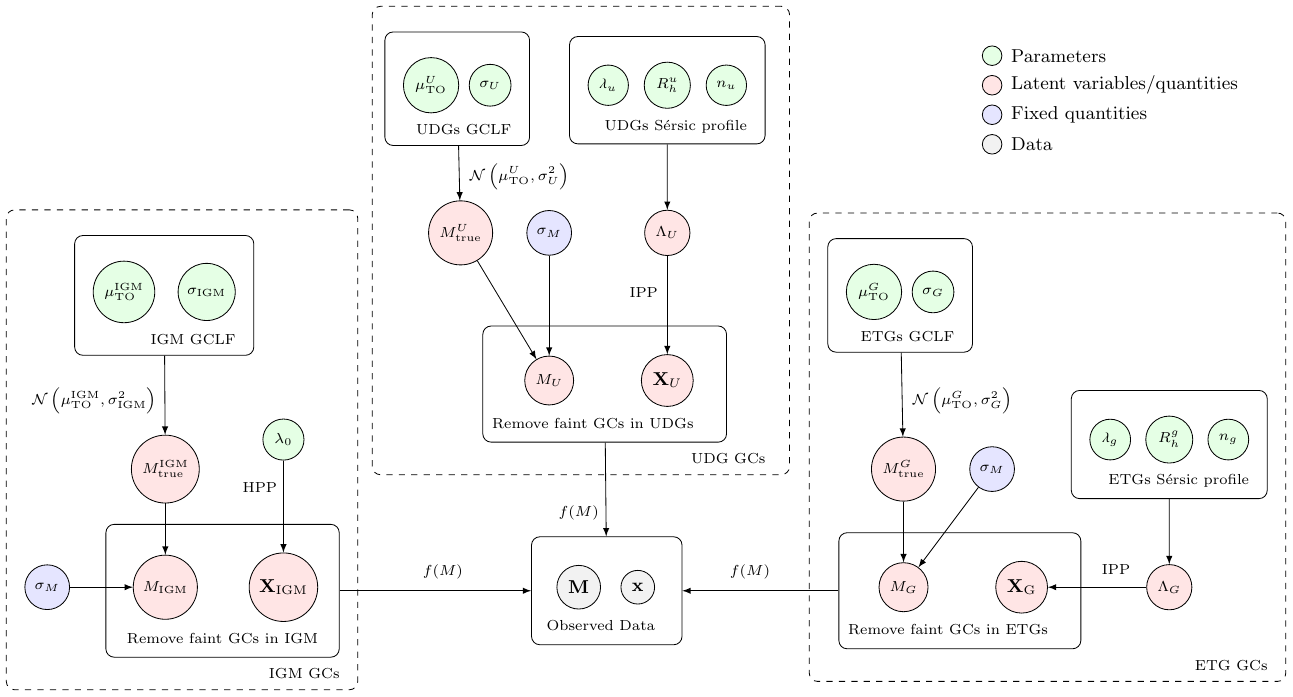}
    \caption{Graphical representation for the data generating process of the observed GC point pattern $\mathbf{x}$ and magnitudes $\mathbf{M}$ under \textsc{Mathpop}. $\mathbf{x}$ and $\mathbf{M}$ come from independent unions of three different branches of data generating processes for GCs in IGM, UDGs, and luminous early type galaxies.}
    \label{fig:graph mod}
\end{figure*}

\subsection{Thinning of Point Process}\label{subsec:ind thin}

The process $\mathbf{X}$ defined in Section \ref{subsec:PP} only produces a GC point pattern $\mathbf{x}$ that we can observe. 
We are actually interested in the \emph{true} GC point process $\mathbf{X}'$ with intensity $\lambda'(s)$ that generates all GCs in a UDG. An advantage of the point process framework is that it can map the true point process $\mathbf{X}'$ to the observed point process $\mathbf{X}$ through an operation called  \emph{thinning}. 

Thinning is the random removal of points in the point process $\mathbf{X}'$. Typically, thinning occurs according to a \emph{thinning probability} $\rho(s) \in [0,1]$ that depends on $s$. If points $\mathbf{s}_i \in \mathbf{X}'$ are thinned \emph{independently} with probability $1 - \rho(\mathbf{s}_i)$, such an operation is called independent-$\rho(s)$ thinning. The resulting thinned point process $\mathbf{X}$ is another point process with intensity $\lambda(s) = \lambda'(s)\rho(s)$. Additionally, if $\mathbf{X}'$ is a Poisson process, then $\mathbf{X}$ is also a Poisson process. Thus, if $\rho(s)$ is known, we can directly infer the property of $\mathbf{X}'$ (what we want) using $\mathbf{X}$ (what we observe).

\subsection{Mark-Dependently Thinned Point Process}\label{sec:MTPP}

In our application, we not only have the observed positions of the GCs, but also have measurements of their magnitudes. Under point process terminology, the magnitudes $\mathbf{M}$ are called the \emph{mark}, meaning they are a characteristic attached to the points. Therefore, a natural way to analyse this type of data is with a \textit{marked point process} \citep{Myllymaki_2009, Myllymaki_2009b}.

We can model the observed GCs with a marked point process $(\mathbf{X}, \mathbf{M}) \subset \mathcal{D}$ where $\mathcal{D} = \mathcal{S}\times\mathcal{M} \subset \mathbb{R}^2\times\mathbb{R}$. The intensity is $\lambda_M(s, m)$, where $(\mathbf{X}, \mathbf{M})$ is obtained from thinning $(\mathbf{X}', \mathbf{M}')$. Thus, the thinning can depend on both GC locations and magnitudes.

The intensity function of $(\mathbf{X}', \mathbf{M}')$ is now $\lambda'_M(s, m) \geq 0$ for $(s,m) \in \mathcal{D}$, and we can write
\[\label{eq:mpp_int}
    \lambda'_M(s,m) = \lambda'(s)\pi'(m\mid s) = \lambda'(s)\pi'(m)
\]
where $\lambda'(s)$ is the intensity of the unmarked point process $\mathbf{X}'$ and $\pi'(m\mid s)$ is the conditional probability density function (p.d.f.) of the mark given the location. For our problem, $\pi'(m\mid s) = \pi'(m)$ is the Globular Cluster Luminosity Function (GCLF). The dependence on $s$ is dropped since GC magnitudes generally do not depend on the GC locations in a galaxy.

Since $\lambda_M(s,m)$ is still the intensity function of a marked point process, we have
\[\label{eqn:mtpp_decomp}
\lambda_M(s,m) = \lambda(s)\pi(m \mid s),
\]
where $\lambda(s)$ is the intensity of $\mathbf{X}$ and $\pi(m \mid s)$ is the p.d.f. of the observed GC magnitudes $\mathbf{M}$ at location $s$. 

We next define a new thinning probability $t(s, m) \in [0,1]$ that can depend on both the location and the magnitude. We assume that the \emph{observed} GC process $(\mathbf{X}, \mathbf{M})$ is obtained from independent $t(s,m)$-thinning of the \emph{true} GC process $(\mathbf{X}', \mathbf{M}')$. An immediate implication is that 
\[\label{eqn:thinned mark dist}
\pi(m \mid s) = \frac{\pi'(m )t(s,m)}{\int \pi'(m)t(s,m)dm} \overset{\triangle}{=} \frac{\pi'(m )t(s,m)}{\Pi'_t(s)}.
\]
The above is effectively applying $t(s,m)$ as a truncation function to $\pi'(m \mid s)$ which produces the magnitude distribution of observed GCs. In our problem, we will assume $t(s,m) = f(m)$, i.e., the thinning of unobserved GCs is caused by the completeness fraction $f(m)$. Therefore, we have
\[
\pi(m \mid s) = \pi(m) = \frac{\pi'(m )f(m)}{\int \pi'(m)f(m)dm} \overset{\triangle}{=} \frac{\pi'(m)f(m)}{\Pi'_f}.
\]

The spatial thinning probability $\rho(s)$ that determines whether a GC at location $s$ is removed or not is then obtained by marginalizing $m$. Hence,
\[\label{eqn:rho}
    \rho(s) =  \int \pi'(m)f(m)dm = \Pi'_f,
\]
and
\[\label{eqn:lambda_t0}
\lambda(s) = \lambda'(s)\Pi'_f.
\]

\subsection{Model for GC Point Process}\label{sec:model}

Following the \textsc{Mathpop} framework in Section \ref{sec:MTPP}, we now present our complete model. For visual aid, Figure \ref{fig:graph mod} shows a graphical representation of our complete model. Since we are using a probabilistic catalog of GCs, we require additional model structures to include the uncertainty of GC candidates. For coherency, we leave the details of these additional structures in Appendix \ref{sec:DGP pp}.

\subsubsection{Modeling the Unmarked GC Point Process}

\begin{figure*}
    \centering
    \includegraphics[width = \textwidth]{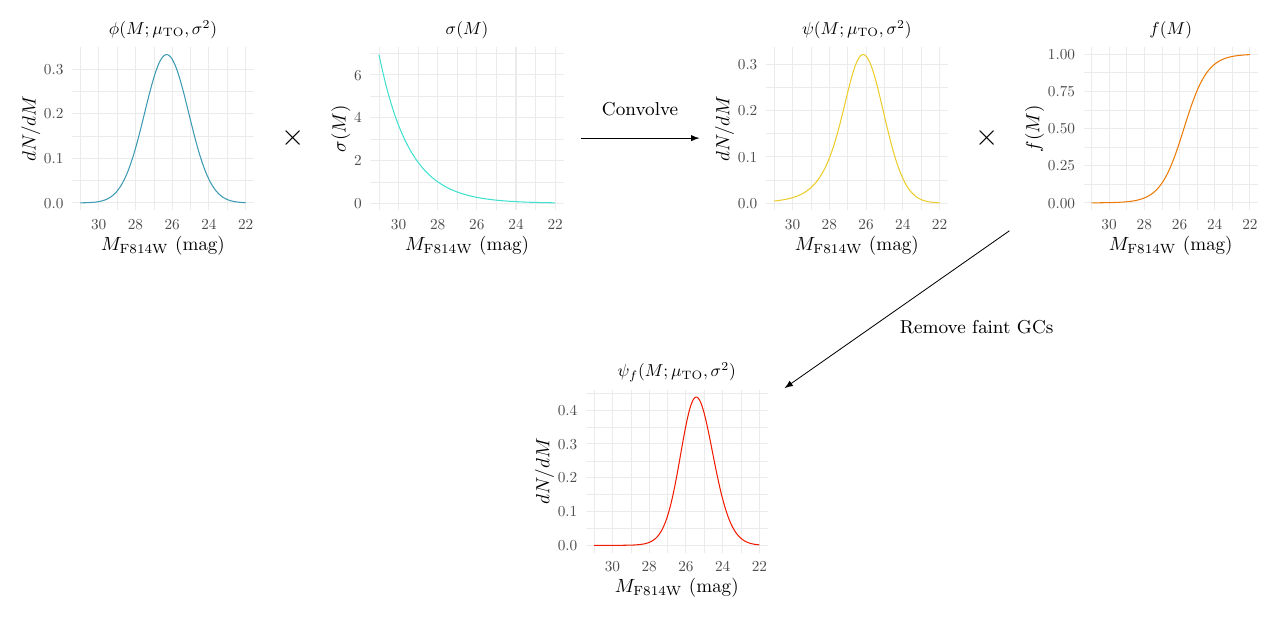}
    \caption{Illustration of our approach to obtain the observed GCLF from the true GCLF. The true GCLF ($\phi(\cdot; \mu_{\text{TO}}, \sigma^2)$; blue) is described by a Gaussian density $\phi(\cdot)$ with TO $\mu_{\text{TO}}$ and dispersion $\sigma$. It is then convolved with the magnitude uncertainty ($\sigma(M)$; turquoise) to obtain the noisy GCLF ($\psi(\cdot)$; yellow). The noisy GCLF is then truncated by the completeness fraction ($f(M)$; orange) to obtain the final observed GCLF ($\psi_f(\cdot)$; red).}
    \label{fig:GCLF}
\end{figure*}


We first focus on the modelling of the unmarked GC point processes $\mathbf{X}$ (observed, thinned) and $\mathbf{X}'$ (true). Let $\Lambda(s) \geq 0$ be the intensity of $\mathbf{X}$, and $\Lambda'(s) \geq 0$ the intensity of $\mathbf{X}'$. 

Note that $\mathbf{X}$ and $\mathbf{X}'$ are more complicated than we have described in Section \ref{sec:MTPP} because the GCs may come from different sub-populations. We assume $\mathbf{X}'$ are independent unions of GC point processes from the following three distinct GC sub-populations:
\begin{description}
 \item[GCs in the Intergalactic Medium (IGM)] Denoted by $\mathbf{X}'_{\mathrm{IGM}}$ with intensity $\Lambda'_{\mathrm{IGM}}(s)$.
    \item[GCs in Luminous Early-Type Galaxies (ETGs)] Denoted by $\mathbf{X}'_G$ with intensity $\Lambda'_G(s)$.
    \item[GCs in UDGs/LSBGs] Denoted by $\mathbf{X}'_U$ with intensity $\Lambda'_U(s)$.
\end{description}
By the independent union assumption, we have
\begin{equation}\label{eqn: superpose}
    \mathbf{X}' = \mathbf{X}'_{\mathrm{IGM}} \cup \mathbf{X}'_G \cup \mathbf{X}'_U,
\end{equation}
and
\begin{equation}\label{eqn: total_int}
    \Lambda'(s) = \Lambda'_{\mathrm{IGM}}(s) + \Lambda'_G(s) + \Lambda'_U(s). 
\end{equation}
The modeling assumptions for GC point processes in each of the above three sub-populations are as follows:
\begin{description}
     \item[IGM] $\mathbf{X}'_{\mathrm{IGM}}$ is a HPP with intensity $\Lambda'_{\mathrm{IGM}}(s) \equiv \lambda_0 > 0$ \citep[see][for a detailed justification]{Harris2020, li2024poisson}.
    \item[ETGs and UDGs/LSBGs] $\mathbf{X}'_G$ and $\mathbf{X}'_U$ are IPPs. We model the GC intensity in ETGs and UDGs/LSBGs using the S\'{e}rsic profile \citep{Harris1991, Wang+2013, Peng2016,  van_Dokkum_2017, Saifollahi2022}:
    \begin{align*}
        &\mathrm{S\acute{e}rsic}(s;\varlambda, R_h, \alpha) = \\
        &\frac{\varlambda b_{\alpha}^{2\alpha}}{2\pi R_h^2 \alpha \Gamma(2\alpha)e}\exp\left(-b_\alpha\left(\frac{r(s)}{R_h}\right)^{1/\alpha}\right),
    \end{align*}
    where
    \begin{align*}
            r^2(s) =& ((s_x - c_x)\cos(\vartheta) - (s_y - c_y)\sin(\vartheta))^2 + \\
    & ((s_x- c_x)\sin(\vartheta)+ (s_y-c_y)\cos(\vartheta))^2/e^2.
    \end{align*}
The S\'{e}rsic profile specified above contains six parameters $(c, \varlambda, \alpha, R_h, \vartheta, e)$ (see Table \ref{tab:parameter table} for their interpretations). $r(s)$ is the ellipsoidal distance from $s = (s_x, s_y) \in \mathcal{S}$ to the known galactic center $c = (c_x, c_y)$ where the ellipsoidal coordinate is determined by the orientation angle $\vartheta$ and the aspect ratio $e$. We assume that $\vartheta$ and $e$ are known, since these parameters of GC systems often align closely with those of the galactic light distribution and are measured with relatively high precision \citep{kisslerpatig1997halo,G_mez_2001, Wang+2013, Saifollahi2022}.
\end{description}

We assume there are $N_G$  ETGs and $N_U$ UDGs in an image $\mathcal{S}$. The main type of ETG we encounter in our data are smaller elliptical galaxies in the Perseus cluster. We assume the GC point processes from all galaxies in image $\mathcal{S}$ are independent. Thus,
$$
 \Lambda'_G(s) = \sum_{g = 1}^{N_G}\mathrm{S\acute{e}rsic}(s;\varlambda_g, R_h^g, \alpha_g),
$$
$$
\Lambda'_U(s) = \sum_{u = 1}^{N_U}\mathrm{S\acute{e}rsic}(s;\varlambda_u, R_h^u, \alpha_u),
$$
and
\[
\Lambda'(s) =& \lambda_0 +  \Lambda'_G(s) + \Lambda'_U(s).
\]
The subscripts $g$ and $u$ in the above represent ETGs and UDGs/LSBGs respectively. We denote the index sets $\{1, \dots, n\}$ and $ \{0, \dots, n\}$ by $[n]$ and $\langle n \rangle$, respectively, for any positive integer $n$. For simplicity, we reindex the intensity functions above with $k \in \langle N \rangle$ where $N = N_G + N_U$. Let $\Lambda'_0(s) = \lambda_0$ and $\Lambda_k'(s), k \in [N]$ be the GC intensity for one of the galaxies, so
$$
\Lambda'(s) = \sum_{k=0}^N\Lambda'_k(s).
$$
Correspondingly, the intensity of the observed/thinned GC point process $\mathbf{X}$ is
$$
\Lambda(s) = \sum_{k=0}^N\Lambda_k(s),
$$
where $\Lambda_k(s) = \Lambda_k'(s)\rho_k(s)$, and $\rho_k(s)$ follows from Eq.~\ref{eqn:rho}. The previous holds since the removal/thinning of faint GCs are independently applied to each of the $N+1$ GC sub-population. Next, we introduce our modeling strategy for the GC magnitudes, and derive $\rho_k(s)$.

\subsubsection{Modeling the Magnitudes (Mark)}
\begin{deluxetable*}{llcll}[t]
\centering
\tablehead{\colhead{Parameter} & \colhead{Meaning} & \colhead{Unit} & \colhead{Prior} & \colhead{Hyper-parameters}}
  \startdata
  $\log(\lambda_0)$ & Log-intensity of $\mathbf{X}'_{\text{IGM}}$ & kpc$^{-2}$ & $\mathcal{N}(\log(\ell_0), \sigma_b^2)$ & $\ell_0$ varies\tablenotemark{$\dag$}, $\sigma_b = 0.4$ \\
  $\varlambda_u$ & Mean number of GCs in UDGs & - & fold-Normal$(\mu_u, \sigma_u^2)$ & $\mu_u = 0$, $\sigma_u = 50$ \\
  $\log(R_h^u)$ & Log-half-number radius of UDG GC systems & kpc & $\mathcal{N}(\log(R_{\mathrm e}^u), \sigma_{r,u}^2)$ & $R_{\mathrm e}^u$ varies\tablenotemark{$\dag$}, $\sigma_{r,u} = 0.5$ \\
  $\log(\alpha_u)$ & Log-S\'{e}rsic index of UDG GC systems & - & $\mathcal{N}(\log(\alpha_u^0), \sigma_{\alpha, u}^2)$ & $\alpha_u^0 = 1$, $\sigma_{\alpha, u} = 0.75$ \\
   $\log(\varlambda_g)$ & Log-mean number of GCs in ETGs & - & $\mathcal{N}(\log(N_{\text{SF}}^g), \sigma_g^2)$ & $N_{\text{SF}}^g$ varies\tablenotemark{$\dag$}, $\sigma_g = 0.25$ \\
  $\log(R_h^g)$ & Log-half-number radius of ETG GC systems & kpc & $\mathcal{N}(\log(3.7R_{\mathrm e}^g), \sigma_{r,g}^2)$ & $R_{\mathrm e}^g$ varies\tablenotemark{$\dag$}, $\sigma_{r,g} = 0.5$ \\
  $\log(\alpha_g)$ & Log-S\'{e}rsic index of ETG GC systems & - & $\mathcal{N}(\log(\alpha_g^0), \sigma_{\alpha, g}^2)$ & $\alpha_g^0 = 0.5$, $\sigma_{\alpha, g} = 0.5$ \\
  $\mu_{\text{TO}}$ & GCLF TO point & mag & $\mathcal{N}(\mu_0, \sigma_{\mu}^2)$ & $\mu_0 = 26.3$, $\sigma_{\mu} = 0.5$\\
  $\log(\sigma)$ & Log-GCLF dispersion & mag & $\mathcal{N}(\log(\sigma_0), \tau_s^2)$ & $\sigma_0 = 1.3$, $\tau_s = 0.25$\\
  \enddata
\tablenotetext{$\dag$}{See Appendix \ref{sec:prior}}
\caption{Prior distributions for parameters in our model.} 
\label{tab:prior table}
\end{deluxetable*}
Let $n$ be the number of GCs observed in an image. We model the conditional distribution of the observed magnitude $\mathbf{M} = \{M_i\}_{i=1}^n$, given $\mathbf{X} = \{\mathbf{s}_i\}_{i=1}^n$ and $\Lambda'$, with a mixture distribution. Each mixture component models the observed GC magnitude distribution in each of the $N+1$ sub-populations:
\[
M_i \mid \mathbf{s}_i,\Lambda', \mu_{\text{TO}}^k, \sigma_k \sim \sum_{k=0}^N\pi_k(\mathbf{s}_i)\mathcal{N}_f(\mu_{\text{TO}}^k, \sigma^2_k).
\]
The mixture probability that the $i$-th GC comes from the $k$-th sub-population is
\[\label{eq:membership prob}
\pi_k(\mathbf{s}_i) = \Lambda'_k(\mathbf{s}_i)/\Lambda'(\mathbf{s}_i).
\]

We denote the observed GC magnitude distribution in the $k$-th sub-population by the random variable $\mathcal{N}_f(\mu_{\text{TO}}^k, \sigma_k^2)$, where $\mu_{\text{TO}}^k$ and $\sigma_k$ are the mean and standard deviation of the true GCLF of the $k$-th sub-population.   By Eq.~\ref{eqn:thinned mark dist}, the p.d.f. of $\mathcal{N}_f(\mu_{\text{TO}}^k, \sigma_k^2)$ is:
\[\label{eqn:psi_f}
    \psi_f(m; \mu_{\text{TO}}^k, \sigma_k^2) &= \frac{\psi(m; \mu_{\text{TO}}^k, \sigma_k^2)f(m)}{\int_{-\infty}^\infty\psi(m; \mu_{\text{TO}}^k, \sigma_k^2)f(m)dm} \\
    &\overset{\triangle}{=} \frac{\psi(m; \mu_{\text{TO}}^k, \sigma_k^2)f(m)}{\Psi_f^k},
\]
where $\psi(m; \mu_{\text{TO}}^k, \sigma_k^2)$ is the noisy GCLF (where the measurement uncertainties of the magnitudes are considered). 

Figure \ref{fig:GCLF} gives a visual representation of our approach to obtain the p.d.f. of $\mathcal{N}_f(\mu_{\text{TO}}^k, \sigma_k^2)$ from the true GCLF. Essentially, $\mathcal{N}_f(\mu_{\text{TO}}^k, \sigma_k^2)$ is the noisy magnitude distribution after thinning by the completeness fraction $f(m)$
(cf. Eq.~\ref{eqn:thinned mark dist}). 

The measured (noisy) magnitude $M$ is generated through the following hierarchical model:
\[\label{eqn:measurement error}
    M \mid M_t &\sim \mathcal{N}(M_t, \sigma_M^2(M_t)), \\
       M_t \mid \mu_{\text{TO}}^k, \sigma_k &\sim \mathcal{N}(\mu_{\text{TO}}^k, \sigma^2_k)
\]
where $M_t$ is the true magnitude and 
$\sigma_M(\cdot)$ is the measurement uncertainty of the magnitude, as described in Section \ref{sec:data}. The p.d.f. of $M$ is thus
\begin{align*}
    &\psi(m; \mu_{\text{TO}}^k, \sigma^2_k) = \\
& \int_{-\infty}^\infty\phi(m; m_t, \sigma_M^2(m_t))\phi(m_t; \mu_{\text{TO}}^k, \sigma^2_k)dm_t,
\end{align*}
where $\phi(m_t; \mu_{\text{TO}}^k, \sigma^2_k)$ is the true GCLF. The term $\phi(m; m_t, \sigma_M^2(m_t))$ is the uncertainty distribution for the observed magnitude $m$, which is also assumed to be Gaussian. $\Psi_f^k$ is then the proportion of GCs that we can observe in the $k$-th sub-population. Computationally, we rely on numerical integration to obtain $\psi(m; \mu_{\text{TO}}^k, \sigma^2_k)$ and $\Psi_f^k$. 

Combining everything from before, our complete hierarchical \textsc{Mathpop} model is
\[\label{eqn:full_model}
    \Lambda'_0(s) =& \lambda_0, \\
    \Lambda'_k(s) =& \mathrm{S\acute{e}rsic}(s;\varlambda_k, R_h^k, \alpha_k), k \geq 1\\
    M_i \mid \mathbf{s}_i,\Lambda', \mu_{\text{TO}}^k, \sigma_k \sim& \sum_{k=0}^N\pi_k(\mathbf{s}_i)\mathcal{N}_f(\mu_{\text{TO}}^k, \sigma^2_k), \ i \in [n] \\
    \pi_k(s) =& \Lambda'_k(s)/\Lambda'(s), \\
    \Lambda_k(s) =& \Lambda'_k(s)\Psi_f^k,\\
    \Lambda(s) =&  \sum_{k = 0}^{N}\Lambda_k(s), \\
    \mathbf{X} \mid \Lambda \sim& \mathrm{IPP}(\Lambda).
\]

\subsection{Prior Distributions and Posterior Sampling}
The prior distributions for our model parameters are in Table \ref{tab:prior table}. The explanation and motivation for choosing these priors are given in Appendix \ref{sec:prior}.

To fit our \textsc{Mathpop} model, we use a Markov chain Monte-Carlo (MCMC) algorithm to sample from the posterior distribution. Specifically, we construct an adaptive Metropolis algorithm \citep{haario2001, roberts2009}. The details of the algorithm are given in the Appendix \ref{sec:inference}. 

We run three independent Markov chains for each image. The chain length is catered to the complexity of the images --- images with more galaxies have more iterations. Posterior convergence diagnostics such as potential scale reduction factors and effective sample size are computed and assessed using the \texttt{R} package \texttt{posterior} \citep{posterior}.

\section{Results}\label{sec:res}

In this section, we present the results and analysis obtained by fitting our \textsc{Mathpop} model to the GC data obtained from the PIPER survey. 

\begin{figure*}[t]
    \centering
    \includegraphics[width = 0.65\linewidth]{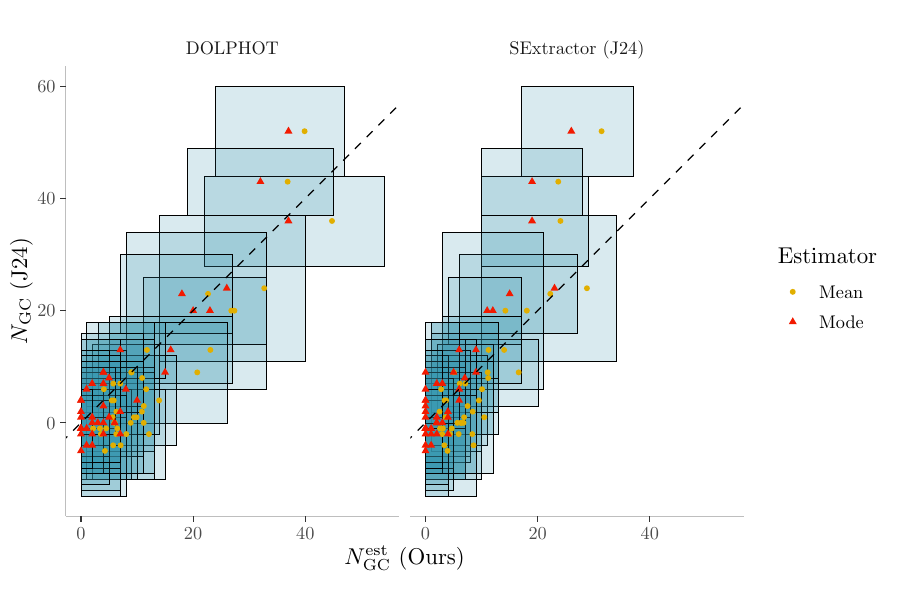}
    \caption{Comparison of the estimated number of GCs around LSBGs. Left: results ($N_{\mathrm{GC}}^{\mathrm{est}}$) obtained from \textsc{MATHPOP} using the L24 catalog (usig point sources from DOLPHOT) vs. the standard approach ($N_{\mathrm{GC}}$ (J24)). Right: results ($N_{\mathrm{GC}}^{\mathrm{est}}$) obtained from \textsc{MATHPOP} using the Prob J24 catalog (J24 point source data, SExtractor) vs. the standard approach ($N_{\mathrm{GC}}$ (J24)). Red triangles are our posterior mode estimates, yellow points are our posterior mean estimates. The blue boxes are the uncertainties of each estimate: the height of the box is $2\times$ the standard error of the estimates from the standard approach; width of the box is the $68\%$ credible intervals from our estimates. Black dashed lines are the $y = x$ reference line.}\label{fig:NGC_est}
    \label{fig:NGC comparison}
\end{figure*}

\subsection{GC Counts}

Figure \ref{fig:NGC_est} compares our \textsc{Mathpop} model estimates of the number of GCs ($N^{\text{est}}_{\text{GC}}$) to the number of GCs estimated by the standard approach ($N_{\text{NG}}$) using the binary GC catalog from J24 (hereafter Binary J24 catalog). The orange circles and the red triangles are the posterior mean and mode from our \textsc{Mathpop} results, respectively. The widths of the blue boxes show the uncertainty in our estimates (68\% credible interval) and the heights show the uncertainty in the J24 estimates (two standard errors). The left panel of Figure \ref{fig:NGC_est} presents the \textsc{Mathpop} results based on the probabilistic GC catalog from the DOLPHOT point sources (hereafter L24 catalog), and the right panel shows the results obtained from the probabilistic GC catalog obtained using the point source list from J24 (hereafter Prob J24 catalog). 

For LSBGs with higher estimates of $N_{\text{GC}}$, our credible intervals are wider than those from the standard approach. The reverse is true for LSBGs with low GC counts. This is expected since the variance of the Poisson distribution increases with the mean. Furthermore, \textsc{Mathpop} fully accounts for all sources of uncertainties. In contrast, the standard approach assumes that $\mu_{\text{TO}}$ is fixed and does not account for its uncertainty. For LSBGs with low $N_{\text{GC}}$, our uncertainty ranges are lower bounded by zero, so our credible intervals are much narrower than those from the standard approach. 

\begin{deluxetable*}{lccrrrcccrc}[t]
\centering
\tablehead{\colhead{ID} & \colhead{$\alpha$} & \colhead{$\delta$} & \colhead{$N_{\mathrm{GC}}^{\mathrm{est}}$} & \colhead{$N_{\mathrm{GC}}^{\mathrm{est}}$} & \colhead{$N_{\mathrm{GC}}$ (J24)} & \colhead{$\mathbb{P}(N_{\mathrm{GC}} = 0)$} & \colhead{$\mu_{\text{TO}}$} & \colhead{$\sigma$}  & \colhead{$R_{h}$} & \colhead{UDG?}\\
    & \colhead{(J2000.0)} & \colhead{(J2000.0)} & \colhead{Mode} & \colhead{Mean}  & \colhead{} & \colhead{} & \colhead{(mag)} & \colhead{(mag)} & \colhead{(kpc)} & \colhead{}}
  \startdata
  R21 & 03:20:29.5 & 41:44:51.0 & $37_{-15}^{+17}$ & $44.76$ & $36 \pm 8$ & $0.000$ & $26.13_{-0.36}^{+0.39}$ & $1.21_{-0.26}^{+0.19}$ & $2.35_{-0.65}^{+0.87}$ & \checkmark \\ 
  R27 & 03:19:43.6 & 41:42:46.8 & $37_{-13}^{+10}$ & $39.88$ & $52 \pm 8$ & $0.000$ & $25.75_{-0.24}^{+0.32}$ & $0.99_{-0.26}^{+0.14}$ & $2.03_{-0.52}^{+0.65}$ & $\times$ \\ 
  R84 & 03:17:24.9 & 41:44:21.5 & $32_{-13}^{+13}$ & $36.85$ & $43 \pm 6$ & $0.000$ & $26.12_{-0.39}^{+0.42}$ & $1.28_{-0.28}^{+0.19}$ & $1.33_{-0.32}^{+0.41}$ & \checkmark \\
  W88 & 03:19:59.1 & 41:18:32.4 & $26_{-12}^{+14}$ & $32.67$ & $24 \pm 13$ & $0.000$ & $25.82_{-0.47}^{+0.43}$ & $1.30_{-0.33}^{+0.23}$ & $3.80_{-1.16}^{+1.65}$ & \checkmark \\
  R60 & 03:19:36.2 & 41:57:26.2 & $23_{-12}^{+10}$ & $27.34$ & $20 \pm 6$ & $0.000$ & $26.01_{-0.43}^{+0.48}$ & $1.28_{-0.32}^{+0.22}$ & $1.05_{-0.34}^{+0.47}$ & $\times$ \\ 
  W89 & 03:20:00.1 & 41:17:05.4 & $20_{-12}^{+13}$ & $26.80$ & $17 \pm 14$ & $0.000$ & $26.29_{-0.38}^{+0.43}$ & $1.12_{-0.32}^{+0.21}$ & $1.53_{-0.60}^{+0.77}$ & \checkmark \\ 
  R16 & 03:18:36.5 & 41:11:32.1 & $18_{-11}^{+9}$ & $22.66$ & $23 \pm 7$ & $0.000$ & $26.11_{-0.`38}^{+0.44}$ & $1.14_{-0.33}^{+0.20}$ & $3.81_{-1.27}^{+1.77}$ & \checkmark \\ 
  R5 & 03:17:34.6 & 41:45:21.6 & $16_{-11}^{+11}$ & $23.07$ & $13 \pm 6$ & $0.000$ & $26.29_{-0.48}^{+0.45}$ & $1.38_{-0.34}^{+0.21}$ & $1.12_{-0.61}^{+0.88}$ & \checkmark \\ 
  W29 & 03:18:23.3 & 41:45:00.6 & $15_{-10}^{+11}$ & $20.73$ & $9 \pm 9$ & $0.003$ & $26.44_{-0.46}^{+0.42}$ & $1.23_{-0.32}^{+0.21}$ & $1.85_{-0.66}^{+0.75}$ & $\times$ \\
  W56 & 03:18:48.1 & 41:14:02.2 & $10_{-7}^{+7}$ & $13.94$ & $4 \pm 8$ & $0.003$ & $26.22_{-0.46}^{+0.46}$ & $1.18_{-0.32}^{+0.21}$ & $0.48_{-0.19}^{+0.33}$ & $\times$ \\
  W79 & 03:18:21.2 & 41:46:15.3 & $8_{-6}^{+6}$ & $11.63$ & $6 \pm 8$  & $0.007$ & $26.27_{-0.44}^{+0.46}$ & $1.16_{-0.35}^{+0.20}$ & $1.23_{-0.55}^{+0.89}$ & \checkmark \\
  R15 & 03:17:03.8 & 41:14:55.0 & $7_{-4}^{+8}$ & $11.76$ & $13 \pm 5$ & $0.008$ & $26.48_{-0.41}^{+0.48}$ & $1.08_{-0.34}^{+0.21}$ & $1.87_{-0.74}^{+1.10}$ & \checkmark \\
  W22 & 03:18:05.4 & 41:27:42.1 & $7_{-5}^{+8}$ & $12.12$ & $-2 \pm 8$ & $0.013$ & $26.33_{-0.47}^{+0.45}$ & $1.21_{-0.35}^{+0.20}$ & $1.39_{-0.52}^{+0.67}$ & $\times$ \\
  W12 & 03:17:36.7 & 41:23:00.9 & $7_{-6}^{+6}$ & $10.86$ & $2 \pm 7$ & $0.012$ & $26.36_{-0.44}^{+0.45}$ & $1.16_{-0.35}^{+0.20}$ & $1.33_{-0.62}^{+0.82}$ & \checkmark \\ 
  W7 (V12) & 03:17:16.0 & 41:20:11.7 & $6_{-5}^{+7}$ & $11.18$ & $0 \pm 10$ & $0.017$ & $26.50_{-0.46}^{+0.45}$ & $1.11_{-0.34}^{+0.20}$ & $0.95_{-0.39}^{+0.45}$ & $\times$\\ 
  W1 & 03:17:00.4 & 41:19:20.6 & $5_{-4}^{+8}$ & $10.90$ & $8 \pm 10$ & $0.020$ & $26.26_{-0.50}^{+0.48}$ & $1.26_{-0.39}^{+0.25}$ & $1.39_{-0.66}^{+1.09}$ & \checkmark \\
  W13 & 03:17:38.2 & 41:31:56.6 & $5_{-5}^{+6}$ & $9.44$ & $1 \pm 7$ & $0.022$ & $26.27_{-0.53}^{+0.46}$ & $1.35_{-0.36}^{+0.23}$ & $1.03_{-0.44}^{+0.59}$ &  $\times$ \\
  W5 & 03:17:10.9 & 41:34:03.6 & $4_{-4}^{+7}$ & $9.01$ & $9 \pm 7$ & $0.036$ & $26.38_{-0.45}^{+0.47}$ & $1.11_{-0.36}^{+0.22}$ & $1.96_{-0.97}^{+1.46}$ & \checkmark \\
  W7 (V14) & 03:17:16.0 & 41:20:11.7 & $4_{-4}^{+10}$ & $8.88$ & $0 \pm 10$ & $0.038$ & $26.34_{-0.50}^{+0.47}$ & $0.92_{-0.62}^{+0.23}$ & $0.94_{-0.42}^{+0.57}$ & $\times$ \\
  W17 & 03:17:44.1 & 41:21:18.7 & $4_{-4}^{+4}$ & $7.00$ & $7 \pm 6$ & $0.032$ & $26.23_{-0.47}^{+0.49}$ & $1.22_{-0.37}^{+0.23}$ & $0.99_{-0.43}^{+0.69}$ & \checkmark \\   
  R23 & 03:19:51.5 & 41:54:35.6 & $4_{-3}^{+7}$ & $8.98$ & $9 \pm 6$ & $0.031$ & $26.37_{-0.50}^{+0.48}$ & $1.21_{-0.39}^{+0.23}$ & $2.14_{-1.02}^{+1.82}$ & \checkmark \\ 
  R20 & 03:20:24.6 & 41:43:28.6 & $4_{-4}^{+5}$ & $8.08$ & $-2 \pm 8$ & $0.039$ & $26.17_{-0.52}^{+0.52}$ & $1.37_{-0.39}^{+0.25}$ & $0.99_{-0.47}^{+0.75}$ & \checkmark \\
  W4 & 03:17:07.1 & 41:22:52.4 & $4_{-4}^{+9}$ & $11.21$ & $3 \pm 12$ & $0.044$ & $26.32_{-0.46}^{+0.52}$ & $0.95_{-0.61}^{+0.24}$ & $3.28_{-1.44}^{+2.08}$ &  $\times$ \\
  \hline
  W8  & 03:17:19.6 & 41:34:32.0 & $3_{-3}^{+4}$ & $6.02$ & $0 \pm 7$ & $0.071$ & $26.44_{-0.50}^{+0.46}$ & $1.15_{-0.39}^{+0.22}$ & $1.08_{-0.48}^{+0.77}$ & \checkmark \\
  W80 & 03:19:39.2 & 41:13:43.5 & $2_{-2}^{+5}$ & $5.76$ & $7 \pm 8$ & $0.078$ & $26.41_{-0.48}^{+0.50}$ & $1.18_{-0.36}^{+0.24}$ & $1.30_{-0.61}^{+0.76}$ &  $\times$ \\ 
  W6 & 03:17:13.3 & 41:22:07.5 & $2_{-2}^{+5}$ & $6.29$ & $-2 \pm 10$ & $0.080$ & $26.33_{-0.50}^{+0.52}$ & $1.15_{-0.73}^{+0.28}$ & $0.62_{-0.31}^{+0.45}$ & $\times$ \\ 
  W59 & 03:18:54.3 & 41:15:28.9 & $2_{-2}^{+5}$ & $5.92$ & $0 \pm 9$  & $0.082$ & $26.30_{-0.52}^{+0.50}$ & $1.27_{-0.38}^{+0.23}$ & $0.60_{-0.29}^{+0.42}$ & $\times$ \\
  W2 & 03:17:03.3 & 41:20:28.7 & $0_{-0}^{+11}$ & $9.93$ & $1 \pm 10$  & $0.082$ & $26.40_{-0.50}^{+0.49}$ & $1.19_{-0.37}^{+0.23}$ & $3.74_{-1.76}^{+2.46}$ & $\times$ \\
  W18 & 03:17:48.4 & 41:18:39.3 & $2_{-2}^{+6}$ & $7.07$ & $-4 \pm 9$  & $0.083$ & $26.36_{-0.53}^{+0.48}$ & $1.22_{-0.39}^{+0.23}$ & $2.13_{-1.05}^{+2.03}$ & \checkmark \\
  W14 & 03:17:39.2 & 41:31:03.5 & $1_{-1}^{+7}$ & $6.46$ & $-1 \pm 7$  & $0.089$ & $26.40_{-0.50}^{+0.50}$ & $1.21_{-0.39}^{+0.21}$ & $1.04_{-0.50}^{+0.78}$ & $\times$ \\
  R117 & 03:18:03.8 & 41:27:08.8 & $2_{-2}^{+5}$ & $5.65$ & $1 \pm 8$  & $0.094$ & $26.30_{-0.49}^{+0.52}$ & $1.31_{-0.39}^{+0.22}$ & $0.65_{-0.31}^{+0.46}$ & $\times$ \\ 
  R79 & 03:18:21.2 & 41:46:15.3 & $1_{-1}^{+7}$ & $5.74$ & $-4 \pm 9$ & $0.102$ & $26.38_{-0.52}^{+0.50}$ & $1.22_{-0.39}^{+0.22}$ & $0.95_{-0.46}^{+0.66}$ & $\times$ \\
  R116 & 03:17:46.0 & 41:30:11.6 & $0_{-0}^{+7}$ & $5.85$ & $4 \pm 8$ & $0.130$ & $26.50_{-0.52}^{+0.49}$ & $1.18_{-0.39}^{+0.22}$ & $1.22_{-0.54}^{+0.88}$ & $\times$ \\
  W28 & 03:18:21.7 & 41:45:27.5 & $0_{-0}^{+7}$ & $6.30$ & $2 \pm 9$  & $0.132$ & $26.48_{-0.51}^{+0.51}$ & $1.17_{-0.38}^{+0.23}$ & $0.65_{-0.35}^{+0.50}$ & $\times$ \\
  W83 & 03:19:47.4 & 41:44:08.8 & $1_{-1}^{+5}$ & $4.07$ & $6 \pm 7$  & $0.134$ & $26.37_{-0.48}^{+0.50}$ & $1.20_{-0.37}^{+0.24}$ & $0.67_{-0.30}^{+0.47}$ & $\times$ \\ 
  W25 & 03:18:15.5 & 41:28:35.3 & $0_{-0}^{+6}$ & $5.49$ & $4 \pm 6$  & $0.167$ & $26.41_{-0.53}^{+0.50}$ & $1.25_{-0.40}^{+0.23}$ & $1.09_{-0.54}^{+0.82}$ & $\times$ \\
  R14 & 03:17:06.1 & 41:13:03.2 & $0_{-0}^{+5}$ & $4.48$ &  $-1 \pm 5$ &$0.178$ & $26.47_{-0.55}^{+0.52}$ & $1.18_{-0.39}^{+0.22}$ & $1.39_{-0.73}^{+1.15}$ & \checkmark \\
  W19 & 03:17:53.1 & 41:19:31.9 & $0_{-0}^{+5}$ & $4.27$ & $-5 \pm 6$  & $0.195$ & $26.46_{-0.52}^{+0.54}$ & $1.21_{-0.41}^{+0.24}$ & $1.94_{-0.93}^{+1.37}$ & \checkmark \\
  W16 & 03:17:41.8 & 41:24:02.0 & $0_{-0}^{+4}$ & $3.54$ & $-2 \pm 7$ & $0.217$ & $26.42_{-0.52}^{+0.49}$ & $1.21_{-0.39}^{+0.24}$ & $1.21_{-0.60}^{+0.78}$ & $\times$ \\ 
  R89 & 03:20:12.8 & 41:44:57.7 & $0_{-0}^{+4}$ & $3.38$ & $-1 \pm 7$  & $0.261$ & $26.46_{-0.54}^{+0.52}$ & $1.21_{-0.41}^{+0.23}$ & $0.53_{-0.26}^{+0.37}$ & $\times$ \\ 
  W84 & 03:19:49.7 & 41:43:42.4 & $0_{-0}^{+2}$ & $2.16$ & $-1 \pm 7$ & $0.335$ & $26.40_{-0.50}^{+0.54}$ & $1.22_{-0.39}^{+0.25}$ & $0.68_{-0.34}^{+0.45}$ & $\times$ \\   
  \enddata
\caption{Posterior predictive summary of GC system properties of $40$ LSBGs in the PIPER survey. $\alpha$ and $\delta$ are the celestial coordinates of the LSBGs in right ascension and declination, respectively. The posterior mode of $N_{\text{GC}}$ together with the $68\%$-highest posterior density interval is given in column 4. Posterior mean of $N_{\text{GC}}$ are given in columns 5. Column 6 contains the $N_{\mathrm{GC}}$ estimates from J24. $\mathbb{P}(N_{\text{Gc}})$ is the probability that an LSBG has no GCs.  $\mu_{\text{TO}}$ and $\sigma$ give the posterior mode of the GCLF TO points and dispersion, respectively. $R_{h}$ is the posterior mode of the half-number radius of GC systems. The last column indicates whether or not an LSBG meets the strict criteria of being a UDG. The horizontal line separates the LSBGs into ones with $\mathbb{P}(N_{\mathrm{GC}} = 0) \leq 0.05$ (above horizontal line), and ones with $\mathbb{P}(N_{\mathrm{GC}} = 0) > 0.05$ (below horizontal line).} 
\label{tab:GC count results}
\end{deluxetable*}

Based on the left panel of Figure \ref{fig:NGC_est} and taking into account the statistical uncertainy, \textsc{Mathpop} results and those from the standard approach generally agree. At first glance, it seems that the posterior mode is a better summary statistic than the posterior mean, especially for LSBGs with lower $N_{\text{GC}}$. This is expected since the posterior distributions of Poisson counts are generally right skewed, and the skewness is even more pronounced for low-count Poisson distributions. For a detailed account on which summary statistic is the best for $N_{\text{GC}}$, we assess the performance of different estimators through simulations in Section \ref{subsec: sim_GC_count}. In short, for low-$N_{\text{GC}}$ LSBGs, we strongly recommend using the posterior mode as the point estimate due to the aforementioned skewness of the posterior distribution.

The right panel of Figure \ref{fig:NGC_est} shows that the results obtained from \textsc{MATHPOP} using the Prob J24 catalog are typically lower than \textsc{MATHPOP} estimates using L24 catalog and those from the standard approach.

An advantage of \textsc{Mathpop} is that, for low-$N_{\text{GC}}$ LSBGs, the estimates and credible intervals do not extend below zero. Moreover, \textsc{Mathpop} provides the posterior probability that an LSBG does not have any GCs ($\mathbb{P}(N_{\text{GC}} = 0)$, Table \ref{tab:GC count results}). As seen in Table \ref{tab:GC count results}, the \textsc{Mathpop} results show that LSBGs with $\mathbb{P}(N_{\text{GC}} = 0) > 5\%$ generally have $N_{\text{GC}}$ estimates significantly above zero, and vice versa. Simulations are carried out in later sections to study how $\mathbb{P}(N_{\text{GC}} = 0)$ behaves.

The detailed posterior summary statistics of GC system properties obtained using L24 catalog, as well as $N_{\text{GC}}$ estimates quoted from the standard approach, are given in Table \ref{tab:GC count results}. The results obtained using Prob J24 catalog are not shown in Table \ref{tab:GC count results} due to space constraints.
 We stress that the point estimates in Table \ref{tab:GC count results} should only be used as a rough guide, as different types of point estimators under Bayesian approach optimize different statistical properties. It is up to the reader to decide which of these properties are preferred. Best practice is to treat the parameters as random variables and to consider their estimated value and uncertainty based on the entire posterior distribution.

Despite the good alignment of estimates from \textsc{Mathpop} and those from the standard approach, the results from Table \ref{tab:GC count results} indicate that for R27 and R84, although not statistically significant, our point estimates of $N_{\text{GC}}$ are much lower than those from the standard approach. We will investigate this discrepancy in the next section.

\subsection{R27 and R84: A Diagnostic}

We investigate the difference in the results from different data and approaches by analyzing R27 and R84 in more detail. Figure \ref{fig:R27&R84} shows the spatial locations of the GC candidates from the three GC catalogs around R27 and R84 within a circular diameter of $15$~kpc, which is the aperture size used by the standard approach to derive $N_{\text{GC}}$ estimates. The GCs are color-coded with their magnitudes in F814W. From left to right panels, the GC candidates, respectively, come from L24 catalog (DOLPHOT sources, left), Binary J24 catalog (SExtractor sources, binary classification, middle), and Prob J24 catalog (SExtractor sources, probabilistic classification, right). The size of the point is proportional to their probability of being a GC ($p(\text{GC})$). For Binary J24 catalog, points have a size of one as they are all fully considered as GCs. For visualization purposes, only points with $p(\text{GC}) > 0.05$ and redder than $0.8$~mag are shown.

\subsubsection{R27}

\begin{figure*}[t]
    \centering
    \includegraphics[width =.9\linewidth]{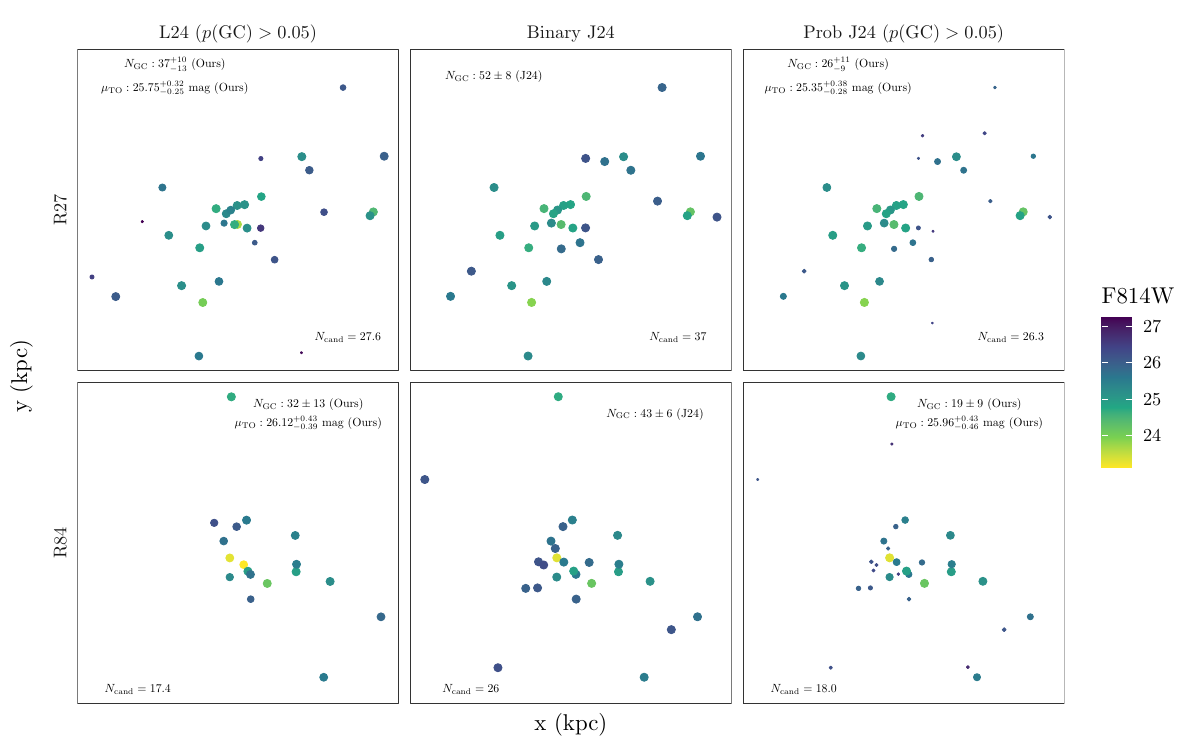}
    \caption{Spatial distribution of GC candidates around the LSBGs (top) R27 and (bottom) R84 within a 15~kpc diameter. For both top and bottom panels: colors are the magnitude of sources in F814W; left panels: GC candidates from L24 catalog; middle panels: GC candidates from Binary J24 catalog; right panels: GC candidates from Prob J24 catalog. The size of points in all panels are proportional to the probability that a source is a GC. For Binary J24 catalog, the size of the point is set to one as a reference, since they are all considered GCs. Estimates of $N_{\text{GC}}$ based on each of the three catalogs are provided for both LSBGs. For L24 catalog and the Prob J24 catalog, the estimated (posterior mode) GCLF TO points ($\mu_{\text{TO}}$) for the two galaxies are also provided. The numbers of GC candidates ($N_{\text{cand}}$; not background-corrected) within the cutouts from each catalog are provided at the bottom of each sub-panel.}
    \label{fig:R27&R84}
\end{figure*}

In the top panel of Figure \ref{fig:R27&R84}, the Binary J24 catalog has $37$ GC candidates around R27 in the cutout, while L24 catalog has $\sim27$\footnote{Computed by summing $p(\text{GC})$ for all sources.} after adjusting for $p(\text{GC})$. After a simple correction of background contaminants in both catalogs (following the same procedure in J24), the numbers of GC candidates ($\sim24$) that can contribute to the final $N_{\text{GC}}$ estimate in R27 are the same in both catalogs. 

The main reason \textsc{Mathpop} point estimate ($37$) is much lower than the standard approach ($52$) for R27 is the treatment of the GCLF. The standard approach assumes a fixed canonical GCLF TO of $\mu_{\text{TO}} = 26.3$~mag, while the posterior mode of the GCLF TO based on \textsc{Mathpop} is half a magnitude brighter than the canonical $\mu_{\text{TO}}$. A simple calculation shows that such a difference in $\mu_{\text{TO}}$ leads to $\sim20\%$ reduction in the final $N_{\text{GC}}$ estimate. Thus, if the GCLF is inferred using the Binary J24 catalog, the point estimate of $N_{\text{GC}}$ in R27 will be reduced to $\sim 42$. As a side note, W88 also has a much brighter inferred $\mu_{\text{TO}}$ (Table \ref{tab:GC count results}) than the canonical $\mu_{\text{TO}}$, but \textsc{Mathpop} estimate is almost the same as that from the standard approach. This is because the image (V11ACS) in which W88 resides is severely affected by background contaminants. In fact, numerous clear imaging artifacts and extended sources are seen in V11ACS from the Binary J24 catalog, which likely resulted in an overestimated background contaminant count under the standard approach and a similar $N_{\text{GC}}$ estimate for W88.

Another minor reason that can further decrease $N_{\text{GC}}$ estimate is the GC membership uncertainty. The standard approach assumes that all GC candidates within the $15$~kpc aperture belong to R27, after accounting for background contaminants. However, simple corrections for background contaminants do not consider the spatial distribution of GCs. It is entirely possible that the GC candidates in the outer region of R27 shown in Figure \ref{fig:R27&R84} actually belong to the IGM. In fact, the average posterior probabilities ($\pi_k(s_i)$, cf. Eq.~\ref{eq:membership prob}) that these outer region GC candidates do belong to R27 are only around $40-60\%$. Such high uncertainties further drives down the final point estimate of $N_{\text{GC}}$. We need to note that the GC membership uncertainty computed by \textsc{Mathpop} is only the best estimate purely based on the spatial distribution of GCs and the S\'{e}rsic profile assumption. True GC membership will not be known without spectroscopy data.


\subsubsection{R84}

For R84, as seen in the bottom panel of Figure \ref{fig:R27&R84}, the inferred $\mu_{\text{TO}}$ from our data is close to the canonical $\mu_{\text{TO}}$, but the number of sources from L24 catalog is fewer. This is in part due to the point source selection criteria between DOLPHOT (L24) and \textsc{SExtractor} (J24). 

Close inspection of the GC candidates around R84 that are in Binary J24 catalog but not in L24 catalog reveals that all are at the boundary of being point sources and extended sources to the human eye, and a few of them are also relatively red. Thus, the true nature of these sources is rather uncertain, and it is difficult to determine whether they are bona-fide GCs or background galaxies. Therefore, the difference in the number of GC candidates considered results in a rather large difference between \textsc{Mathpop} point estimate and that from the standard approach.

\begin{figure*}[t]
    \centering
    \includegraphics[width =0.9\linewidth]{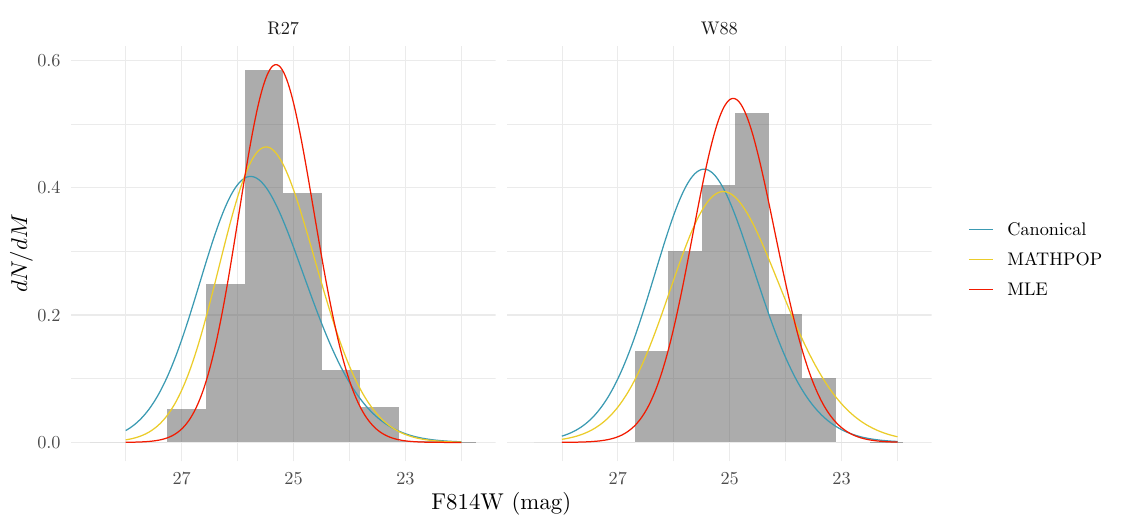}
    \caption{Observed GCLFs ($\psi_f(m; \mu_{\text{TO}}, \sigma^2)$) obtained from different approaches against the observed background-corrected GC magnitude data for R27 and W88. Colored lines are the observed GCLFs based on different assumptions/models. Histograms are the magnitudes of observed GC candidates within $6$~kpc of the galactic center of each galaxy from the L24 catalog. For visualization purposes, uncertainties of estimates are omitted.}
    \label{fig:GCLF_R27_W88}
\end{figure*}

\subsubsection{Prob J24 Catalog}

From the right panels of Figure \ref{fig:R27&R84}, it is clear why the results obtained by using Prob J24 catalog are significantly lower than the other two: fainter sources have lower probability of being GCs within the Prob J24 catalog. This is caused by the significant contaminant population at the faint levels in the J24 point source list, which effectively reduces our ability to separate GCs from contaminants. The reduction of our discriminatory ability of faint sources then leads to a double whammy: sources that are true GCs at faint levels have lower impact in the model, which reduces the count estimates and also causes the inferred $\mu_{\text{TO}}$ to be much brighter. The brighter inferred $\mu_{\text{TO}}$ then further drives the final $N_{\text{GC}}$ estimate downward.


Based on the analysis in Figure \ref{fig:R27&R84}, the true $N_{\text{GC}}$ for R27 and R84 should be upper bounded by estimates from the standard approach and lower bounded by the results from Prob J24 catalog. The results obtained using L24 catalog may be closest to the true value, since our point source catalog (L24) is carefully pruned and cleaned to ensure that the least number of contaminants are present. Furthermore, \textsc{Mathpop} fully addresses all sources of uncertainties, such as inferring $\mu_{\text{TO}}$ from the data and the GC membership uncertainty. 

As a final note, the construction of point source and GC catalogs has immense impact on the final results. Therefore, we recommend the use of `DOLPHOT+\textsc{Mathpop}' for GC catalog construction and inferring the GC counts. Point sources catalogs based on \textsc{SExtractor}, as seen previously, can contain so many contaminants that it is unlikely to produce reliable GC count estimates.

\subsection{GCLF}\label{sec:GCLF}

\begin{figure*}[t]
    \centering
    \subfigure[$N_{\text{GC}}$ estimators performance]{\includegraphics[width = 0.48\linewidth]{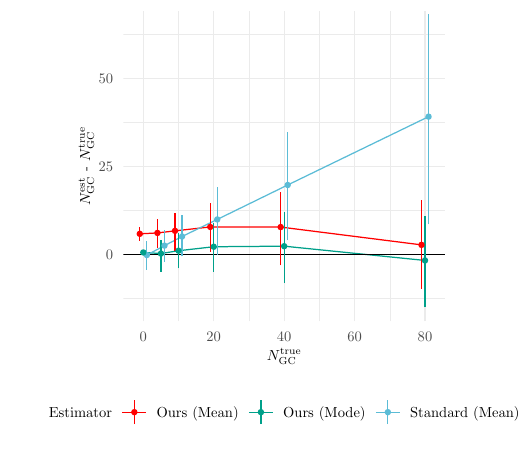}
    \label{fig:Ntrue_vs_Nsim_ours}}
    \subfigure[Standard approach performance under different $\mu_{\text{TO}}$]{\includegraphics[width = 0.48\linewidth]{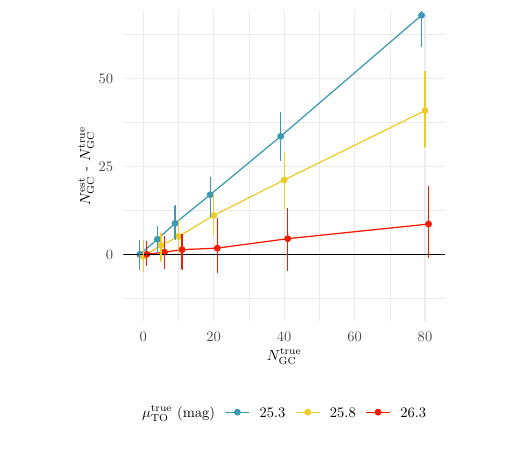}\label{fig:Ntrue_vs_Nmode_mu_stand}} \\
    \subfigure[Posterior mode of $N_{\text{GC}}$ vs. Posterior mode of $\mu_{\text{TO}}$ from \textsc{Mathpop}]
    {\includegraphics[width = 0.75\linewidth]{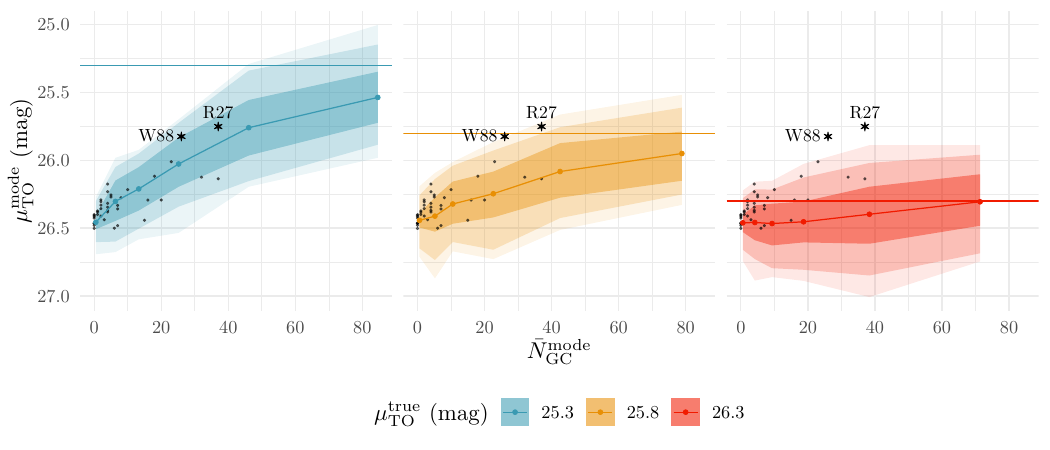}\label{fig:Nmode_vs_mu_mode}}
    \caption{Results from simulation: (a) Point estimates performance (difference between estimates ($N_{\text{GC}}^{\text{est}}$) and true values ($N_{\text{GC}}^{\text{true}}$)) based on $450$ simulated data for each $N_{\text{GC}}^{\text{true}}$; color represents different estimators. (b) Point estimates performance for $N_{\text{GC}}^{\text{true}}$ from the standard approach for different values of $\mu_{\text{TO}}^{\text{true}}$; color represents $\mu_{\text{TO}}^{\text{true}}$. In both (a) and (b), error bars are the central $68\%$ quantile intervals based on the simulations; a trend that is flat and close to $y=0$ line indicates better performance. (c) Average posterior mode estimates of the GC counts ($\bar{N}_{\text{GC}}^{\text{mode}}$) vs. posterior mode of the GCLF TO ($\mu_{\text{TO}}^{\text{mode}}$). The colored points are the respective average posterior mode estimates while the error bands (with decreasing transparency) are the $68, 95$, and $99\%$ central quantile bands for $\mu_{\text{TO}}^{\text{mode}}$. The horizontal lines represent the true values of $\mu_{\text{TO}}$ we used to simulate the GCLF for our UDG. The gray points are the posterior mode estimates of all the LSBGs obtained using L24 catalog. W88 and R27 are marked by black stars.}
\end{figure*}

From Table \ref{tab:GC count results}, the inferred $\mu_{\text{TO}}$ for most LSBGs considered here are close to the canonical $\mu_{\text{TO}} = 26.3$~mag. However, the results of R27 and W88 in Table \ref{tab:GC count results} indicate that they have GCLFs weighted more toward the luminous end, with posterior mode estimates of $\mu_{\text{TO}}$ being more than half a magnitude brighter than the canonical value for both LSBGs. We computed the Bayes' factor \citep[BF;][]{Kass_1995} to determine the amount of evidence on whether $\mu_{\text{TO}}$ is indeed brighter than the canonical value for the two LSBGs using the Savage-Dickey method \citep[\texttt{bayestestR};][BF $>1$ means there is evidence supporting the claim that $\mu_\text{TO}$ is brighter than the canonical value]{Makowski2019}. The resulting BF values are $ 4.5$ and $ 2.5$ for R27 and W88, respectively. 

In addition, as a post-hoc check and crude estimate, we use the GC candidates within $6$~kpc of either LSBG from the L24 catalog, and apply a simple background-correction as in J24 to conduct maximum likelihood estimations (MLE) for R27 and W88 of their respective $\mu_{\text{TO}}$ with completeness fraction considered.

The fitted MLEs of $\mu_{\text{TO}}$ for R27 and W88 are $\mu_{\text{TO}}^{\text{R27}} = 25.42 \pm 0.17$~mag, $\sigma^{\text{R27}} = 0.72 \pm 0.12$~mag and $\mu_{\text{TO}}^{\text{W88}} = 25.17 \pm 0.31$~mag, $\sigma^{\text{W88}} = 0.83 \pm 0.21$~mag. The GCLF TO points for the two LSBGs are respectively $\sim0.9$~mag and $\sim 1.1$~mag brighter than the canonical $\mu_{\text{TO}}$ of $26.3$~mag. The GCLF dispersions are also much smaller than the canonical value of $1 - 1.2$~mag. The $p$-values under the null hypothesis that their $\mu_{\text{TO}}$ is equal to the canonical one are $\sim 10^{-8}$ and $\sim 10^{-5}$, respectively. Based on both the BF values and the MLE results, we thus have significant evidence that these two LSBGs have GCLFs that are significantly more weighted toward the luminous end.

Figure \ref{fig:GCLF_R27_W88} shows the observed GCLFs for R27 and W88 from different approaches against the magnitude distributions of GC candidates within $6$~kpc from the galactic center of both LSBGs. The histograms of GC magnitudes are obtained by resampling the magnitudes of sources (background-corrected) within $6$~kpc from the galactic center of both LSBGs for $500$ times according to the GC candidate probabilities from L24 catalog. For better visualization, uncertainties of estimates are omitted in the figure. The observed GCLFs are computed based on Eq.~\ref{eqn:psi_f}, where the values of $\mu_{\text{TO}}$ and $\sigma$ are set to the posterior mode from \textsc{Mathpop}, the previously obtained MLE, or the canonical value ($\mu_{\text{TO}} = 26.3$ mag, $\sigma = 1.2$~mag). It is clear that for R27 and W88, the canonical GCLF does not fit well to the observed data at all. The GCLFs obtained from the MLE approach closely align with the data as expected. The results from \textsc{Mathpop} are in between the other two as the results from the Bayesian approach are a combination of information from the data and the prior distributions, with the latter based on the canonical values.

The observed top-heavy GCLFs of R27 and W88 put them in a similar category to NGC~1052 DF2 and DF4 \citep{Shen_2021}. DF2 and DF4 have a double-peaked GCLF where there is a major peak that is $\sim1.5$~mag brighter than the canonical TO point and a minor peak at the canonical TO point. However, it is unclear whether R27 and W88 are indeed of the same population and having similar formation scenario as DF2 and DF4, since DF2 and DF4 are much closer and the photometry is much more complete. 




\subsection{Simulated Data}

In this section, we conduct a simulation analysis for our proposed model. Due to computational constraints, we simulate only the point patterns of GCs instead of other possible type of point sources such as foreground stars or background galaxies. Thus, every simulated source is considered a true GC, so the simulation results directly indicate the performance of our proposed model.

For simplicity, we only vary $N_{\text{GC}}$ and $\mu_{\text{TO}}$ of a simulated UDG GC system. We consider an observation window of $76\times76$~kpc$^2$ --- the same as the field of view of ACS images in the PIPER survey. We first simulate the IGM GC locations using an HPP. $N_{\text{GC}}$ from IGM is based on the IGM GC intensity estimated from V10WFC3. We then simulate the UDG GC system from a S\'{e}rsic profile and randomly place it in the field. The GC system has half-number radius $R_h = 2$~kpc, S\'{e}rsic index $\alpha = 2$, aspect ratio $e = 1$ and orientation angle $\vartheta = \pi/4$. $N_{\text{GC}}$ takes a value in $\{0, 5, 10, 20, 40, 80\}$. After the GC point pattern is obtained, we simulate a magnitude for each GC. For GCs in the IGM, they have a canonical GCLF with $\mu_{\text{TO}}=26.3$~mag and dispersion of $1.2$~mag. For GCs in the UDG, they have a GCLF with $\mu_{\text{TO}} = \{25.3, 25.8, 26.3\}$~mag and dispersion of $1.0$~mag. We then jitter these ``true" magnitudes with simulated measurement uncertainties and remove faint GCs using the completeness fraction based on the AST results using DOLPHOT for ACS images. Our model is then fitted to the resulting simulated GC point pattern and magnitude data. For comparison, we also apply the standard approach to the simulated data. Under the standard approach, the simulated data have measurement uncertainties, the completeness fraction follows that from J24, and GCs fainter than $26.3$~mag are removed.

We repeat the above process $150$ times for each parameter configuration, resulting in a total of $6\times 3\times 150 = 2700$ different simulated data. The results are provided and analyzed below.

\subsubsection{GC counts}\label{subsec: sim_GC_count}

Figure \ref{fig:Ntrue_vs_Nsim_ours} shows the performance of $N_{\text{GC}}$ estimates from \textsc{Mathpop} and the standard approach based on the simulated data. On average, the posterior modes from our model nicely match the true $N_{\text{GC}}$. Additionally, the posterior mode is a much better point estimate for $N_{\text{GC}}$ than the posterior mean. As mentioned, this is due to the highly right-skewed nature of the posterior distribution of $N_{\text{GC}}$. For UDGs with low $N_{\text{GC}}$, the posterior mean, as a point estimator, completely overestimates $N_{\text{GC}}$. Therefore, Figure \ref{fig:Ntrue_vs_Nsim_ours} is a definitive showcase that the posterior mode should be used as a point estimator for $N_{\text{GC}}$.

On the other hand, Figure \ref{fig:Ntrue_vs_Nsim_ours} also demonstrates that the standard approach severely overestimates $N_{\text{GC}}$, which is caused by the assumption on GCLF. As seen in Figure \ref{fig:Ntrue_vs_Nmode_mu_stand}, the standard approach is only accurate when the true $\mu_{\text{TO}}$ is the canonical value of $26.3$~mag, while for brighter $\mu_{\text{TO}}$, the standard approach produces completely inaccurate estimates. Since the standard approach assumes all UDGs have the canonical GCLF, the final $N_{\text{GC}}$ estimates are overestimates if the true $\mu_{\text{TO}}$ is brighter than the canonical one.


On another note, we have also tested the effect of the GC system size under the standard approach. We found the assumption that a counting aperture with diameter of $15$~kpc contains $90\%$ (cf. J24) of all GCs in UDGs only works well for a GC system with $R_h \approx 2$~kpc. A deviance of $1$~kpc from $R_h \approx 2$~kpc leads to notable biases of $N_{\text{GC}}$. However, the effect is not as significant as that from inaccurate assumptions of the GCLF as observed in Figure \ref{fig:Ntrue_vs_Nmode_mu_stand}. We did not investigate the effect of changing $R_h$ under \textsc{Mathpop} due to computational constraints, but given that we are also inferring $R_h$ from data, \textsc{Mathpop} should also work well with differing $R_h$.

We also used the simulation results to investigate the relationship between the probability that a UDG has no GC ($\mathbb{P}(N_{\text{GC}} = 0)$) and the true value of $N_{\text{GC}}$ based on our simulations. We found that the previously selected cutoff probability value of $5\%$ corresponds well to when we consider a UDG to have no GC: for $\mathbb{P}(N_{\text{GC}} = 0) > 5\%$, it roughly coincides with the true $N_{\text{GC}}$ being around $1\sim2$. For such a low value, it is effectively not possible for any method to distinguish them from being zero or positive.

\subsubsection{GC counts versus GCLF}\label{sec:NGC vs TO}

Figure \ref{fig:Nmode_vs_mu_mode} shows the relationship between the average posterior mode of $N_{\text{GC}}$ ($\bar{N}_{\text{GC}}^{\text{mode}}$) and the posterior mode distributions of $\mu_{\text{TO}}$ ($\mu_{\text{TO}}^{\text{mode}}$) based on $150$ simulations for each parameter configuration of $N_{\text{GC}}$ and $\mu_{\text{TO}}$. We see that the inferred $\mu_{\text{TO}}^{\text{mode}}$ becomes brighter as $N_{\text{GC}}$ increases, but this is superfluous. Since the prior distribution of $\mu_{\text{TO}}$ is $\mathcal{N}(26.3, 0.5^2)$, and for low-$N_{\text{GC}}$, the lack of data causes the posterior to be similar to the prior. As $N_{\text{GC}}$ increases, $\mu_{\text{TO}}^{\text{mode}}$ gradually approaches the true $\mu_{\text{TO}}$ since more information becomes available.

Furthermore, for low-$N_{\text{GC}}$, $\mu_{\text{TO}}^{\text{mode}}$ is lower than the mean of the prior distribution at $26.3$~mag even if the true $\mu_{\text{TO}}$ is $26.3$~mag. This is because the model has to differentiate whether there are indeed very few GCs or whether the actual $\mu_{\text{TO}}$ is so faint that most of the GCs are unobserved. Balancing these two options leads to, on average, fainter $\mu_{\text{TO}}^{\text{mode}}$ than its prior mean when $N_{\text{GC}}$ is low. 

Although the previously observed relationships in Figure \ref{fig:Nmode_vs_mu_mode} are not physical, they do tell us, for a given $N_{\text{GC}}^{\text{mode}}$ and a true $\mu_{\text{TO}}$, what the possible value of $\mu_{\text{TO}}^{\text{mode}}$ may be. We have plotted in Figure \ref{fig:Nmode_vs_mu_mode} the posterior mode estimates (gray points) obtained by our approach of all LSBGs considered in this paper. The relationship between $N_{\text{GC}}^{\text{mode}}$ and $\mu_{\text{TO}}^{\text{mode}}$ for most LSBGs roughly follows the one when the true $\mu_{\text{TO}}$ is the canonical TO point (red points and confidence bands). 

However, for LSBGs R27 and W88, their posterior mode estimates clearly do not obey the relationship observed for the red points and confidence bands. Given their $N_{\text{GC}}^{\text{mode}}$ values, it is statistically nearly impossible to obtain their respective $\mu_{\text{TO}}^{\text{mode}}$ if their true $\mu_{\text{TO}} = 26.3$~mag. In fact, for UDGs with $N_{\text{GC}}^{\text{mode}}$ similar to R27 and W88, a true $\mu_{\text{TO}}=25.3$~mag, which is a whole magnitude brighter than the canonical TO, is required to have a probable chance of obtaining a $\mu_{\text{TO}}^{\text{mode}}$ similar to R27 and W88. This observation also coincides with the MLEs obtained previously in Section \ref{sec:GCLF}. Even though the MLEs are only crude estimates, the simulation studies here do suggest that the MLEs should be close to the true values. Therefore, we have sufficient evidence from both real data and simulations that the GCLFs of R27 and W88 are indeed much more weighted toward the luminous end than the canonical GCLF. 

The analyses based on our simulation studies demonstrate the accuracy and validity of our approach. Compared to the standard method, our method resolves the issue induced by the variations of the GCLF, which, we show here, is potentially the main driving factor in the massive discrepancy in $N_{\text{GC}}$ estimates of UDGs reported in previous studies. Therefore, it is highly important to infer the GCLF from data rather than making assumptions about it. Otherwise, not only $N_{\text{GC}}$ estimates of UDGs may be inaccurate, important new discovery about UDGs may also be missed.  

\section{Discussion and Conclusions}\label{sec:conclusion}

In this paper, we have introduced a hierarchical Bayesian mark-dependently thinned point process (\textsc{Mathpop}) model to infer the GC counts in UDGs and LSBGs in general. Our method has several notable advantages over previous ones.

Firstly, via our approach, we were able to identify two LSBGs (R27 and W88) with GCLFs significantly more weighted to the luminous end than the canonical GCLF, while it is unlikely that the abnormality of these two UDGs would have been be discovered under the standard approach. These two new discoveries add toward the samples of UDGs that have abnormal GCLFs and may bring about new understanding of galaxy and GC formation theory. The new discoveries are possible since our approach jointly infers the spatial and magnitude distributions of GCs through a marked point process framework. Under this approach, we not only accurately infer and quantify the uncertainties of GC counts, we can also quantify the GCLF of UDGs in a coherent way without need of making assumptions. 

Secondly, under a hierarchical Bayesian point process framework, various uncertainties such as photometric uncertainties and GC membership uncertainties can be fully addressed.

Thirdly, our approach no longer produces GC count estimates or confidence intervals that extend to the negative range. Instead, through simulation studies, we show that the point estimate using the posterior mode under our approach can accurately infer cases of low GC count UDGs. Moreover, our method provides the posterior probability that a UDG has no GC so that proper uncertainty quantification for low GC count UDGs is also available.

Lastly, we provide methods to incorporate the uncertainty associated with GC candidate selection into the entire model fitting procedure while previous methods largely ignored such uncertainties.

While our approach exhibits significant advantages over the standard approach, one main challenge is computation time. Our current MCMC algorithm takes roughly half an hour to run $100$~K iterations for data that contains one UDG. For data with more UDGs, the algorithm can take up to hours. Due to the complexity of our model likelihood, much more efficient gradient-based MCMC algorithms such as Hamiltonian Monte-Carlo \citep{neal2011mcmc} or No-U-Turn sampler \citep{hoffman2014no} cannot be used. We are currently exploring the possibility of using approximate Bayesian inference with surrogate likelihood or posterior as proposed by \citep{li2024bayesian}, which can facilitate fast inference but also provide proper uncertainty quantification.

Another potential issue in this work is that we have not considered realistic simulation of UDGs and their GC systems due to computational constraints. If more computational resources are permitted, more sophisticated software such as \texttt{ARTPOP} \citep{greco2022artpop} can be used to inject highly realistic simulations of UDGs and their GC populations into real images. Moreover, the uncertainty associated with selection of GC candidates can also be assessed.

Additionally, the problematic results obtained by the probabilistic GC catalog from the J24 point source list should be addressed. Although this is not an inherent problem of our \textsc{Mathpop} model, it is an integrated part of our entire Bayesian workflow and the quality of data is ultimately the biggest determining factor on the accuracy of the final estimates. However, with only color-magnitude data, it is impossible to confidently separate GCs from contaminants at faint levels for a highly contaminated point source list such as the one in J24. To address this issue, information on how likely a detected source is indeed a point source can be a crucial factor that boosts our ability to separate GCs from contaminants at faint levels. 

Overall, this work is an initial step at developing an accurate and robust method to infer and quantify the uncertainty of GC counts in UDGs/LSBGs. There are various directions for future research:

Firstly, we plan to speed up our inference algorithm and apply our method to much larger samples of UDGs/LSBGs from various other catalogs.

Secondly, we will consider extensive simulation studies using the \texttt{ARTPOP} software \citep{greco2022artpop} to generate realistic UDGs and their GC systems and inject them into real astronomical images. Based on the simulated images, we will conduct the standard data reduction and GC selection step so that all uncertainties associated with data acquisition can be calibrated and assessed under our approach.

Last but not least, we also intend to develop a probabilistic GC classification method where the likelihood that a source is indeed a point source is incorporated. The development of such a method may not only be useful for our problems but can be crucial for various studies relying on catalogs of extra-galactic GCs.

\section*{Author Contribution}
D.L. contributed to the majority of this work. D.L. conceived the initial idea and proposed the model in the paper. D.L. also wrote majority of the paper and the entire code for model fitting, as well as all the code and analysis to produce all of the figures.

G.M.E. was the PI and co-supervised D.L. G.M.E. provided expertise on Bayesian analysis and astrostatsitics. G.M.E. contributed and edited all sections of the paper.

P.E.B. co-supervised D.L. P.B. also provided expertise on statistics and point process models, edited and contributed to Section \ref{sec:method} and Appendix \ref{sec:DGP pp}.

W.E.H. contributed significantly to Section \ref{sec:data} and obtained the initial point source list through DOLPHOT. W.H. also provided expertise on GCs and edited various sections.

R.G.A. co-supervised D.L., and provided expertise on UDGs/LSBGs. R.A. also edited and contributed to Section \ref{sec:intro} and \ref{sec:res}. 

P.V.D. contributed to Section \ref{sec:res} and provided crucial help in paper structure and software construction.

S.R.J. contributed to Section \ref{sec:data}, obtained and provided point source list from J24. S.J. also provided expertise on UDGs/LSBGs.

S.C.B. edited and contributed to all sections.

S.D. provided various edits and comments to all sections.

A.J.R. provided various edits and comments to all sections.

J.S. provided crucial help to form the initial idea of the paper and provided expertise on astrostatistics.


\section*{Acknowledgments}
Most of the computations in this work were performed using the high performance computing cluster at the Digital Research Alliance of Canada. D.L. would like to acknowledge funding support from Canadian Statistical Sciences Institute and Data Sciences Institute at the University of Toronto through grant number DSI-DSFY2R1P23. W.E.H. would like to acknowledge funding support from NSERC. S.C.B. would like to acknowledge funding support from the Data Sciences Institute at the University of Toronto through grant number DSI-DSFY3R1P24. A.J.R. was supported by National Science Foundation grant AST-2308390. 

\section*{Data}
All the {\it HST} raw imaging data used in this paper can be found in MAST: \dataset[10.17909/t87p-g529]{http://dx.doi.org/10.17909/t87p-g529}.

%

\vspace{5mm}
\facilities{
\textit{HST}
}


\software{
DOLPHOT, \texttt{R}, \texttt{Python}
}



\appendix

\section{Artificial Star Test Results}\label{appdx: AST}

Table \ref{tab:f_err_est} gives the estimated completeness fraction and the measurement uncertainties in both filters based on our AST. 

\begin{table}[t]
   \centering
  \begin{tabular}{l*{8}{c}}
  \hline
  \hline
     Filter (Camera) & $a$ & $\pm a$ & $m_0$ & $\pm m_0$  & $\beta_0$ & $\pm \beta_0$ & $\beta_1$ & $\pm \beta_1$\\
    \hline
    F814W (ACS) & 1.50 & 0.033 & 25.75 & 0.017 & 0.0884 & 0.004 & 0.645 & 0.015\\
    F475W (ACS) & 1.66 & 0.031 & 26.93 &  0.013 & 0.078 & 0.003 & 0.699 &  0.017 \\
    F814W (WFC3) & 1.57 & 0.048 & 26.52 & 0.022 & 0.0977 & 0.003 & 0.613 & 0.011 \\
    F475W (WFC3) & 1.23 & 0.040 & 28.02 & 0.030 & 0.0544 & 0.002 & 0.652 & 0.019\\
    \hline
\end{tabular}
\caption{Estimates of the parameters for the completeness fraction ($a, m_0$) and the measurement uncertainty function ($\beta_0, \beta_1$), and their uncertainties obtained by artificial fake star tests.}
\label{tab:f_err_est}
\end{table}

\section{Probabilistic Classification of Globular Clusters}\label{sec:GC_class}

\subsection{Our Data}\label{Appx A subsec: our data}
For our point source data obtained by DOLPHOT, we apply a multivariate nonparametric finite mixture model to obtain the probabilistic GC catalog using the method of \cite{Benaglia_2009, CHAUVEAU20161}. The method is available in the \texttt{R} package \texttt{mixtools} via the function \texttt{mvnpEM}. This method requires the least amount of information provided to the algorithm and allows the data itself to facilitate clustering. Specifically, we run the algorithm on $500$ simulated color-magnitude distributions, where each simulation is obtained by jittering the color and magnitude of individual point sources using their respective measurement uncertainties. The algorithm is run with the assumption that there are three distinct clusters in the color-magnitude data, where they correspond to (1) the blue star in the lower left corner of the CMD in the left panel of Figure \ref{fig:GC_prob}, (2) the typical GC candidate region, and (3) the Milky Way foreground stars in the upper right region of the CMD in the left panel of Figure \ref{fig:GC_prob}. The probability that each source belongs to the cluster representing the GC candidate region is computed. 

\subsection{J24 Data}\label{Appx A subsec: J24 data}
\begin{table}[t]
\centering %
\resizebox{\linewidth}{!}{%
\begin{tabular}{lclccc}
   \hline\hline
   Component & Mixture Weights &  Variables & Sub-component Weights & Parameters & Distribution \\
   \hline
   \multirow{3}{*}{GCs} & & Magnitude in F814W (mag) &  - &  $\theta_{\text{GC}}^M = (\mu_{\text{TO}}, \sigma_{\text{GC}})$ & $\mathcal{N}\left(\mu_{\text{TO}}, \sigma_{\text{GC}}^2\right)$ \\
   & $w$ & Red GCs in F475W $-$ F814W (mag) & $w_r$ & $\theta_{\text{GC,r}}^C = \left(\mu_{\text{GC,r}}^C, \sigma_{\text{GC,r}}^C\right)$ & $\mathcal{N}\left(\mu_{\text{GC,r}}^C, (\sigma_{\text{GC,r}}^C)^2\right)$ \\
   & & Blue GCs in F475W $-$ F814W (mag) & $1-w_r$ & $\theta_{\text{GC,b}}^C = \left(\mu_{\text{GC,b}}^C, \sigma_{\text{GC,b}}^C\right)$ & $\mathcal{N}\left(\mu_{\text{GC,b}}^C, (\sigma_{\text{GC,b}}^C)^2\right)$ \\
   \hline
   \multirow{2}{*}{Contaminants} &  \multirow{2}{*}{$1-w$} & Magnitude in F814W (mag)  & -  & $\theta_{\text{cont}} = (\mu_{\text{cont}}, \sigma_{\text{cont}})$ & $\mathcal{N}\left(\mu_{\text{cont}}, \sigma_{\text{cont}}^2\right)$ \\
   & & Color in F475W $-$ F814W (mag) & - & - & $\mathrm{Unif}\left(0.8, 2.4\right)$ \\
  \hline
  \end{tabular}%
}%
\caption{Mixture model components and parameters used for clustering point sources from J24 into GCs and contaminants.}
\label{tab:parameter mixture}
\end{table}

For the point source list in J24, the contaminating sources at faint levels are so abundant that the previous nonparametric finite mixture model has difficulties at identifying GC candidates near faint levels. Even within the binary J24 GC catalog, the contaminant populations are still abundant. As a definitive test on the existence of a contaminant population, we follow J24 and assume that sources with F814W $< 26.3$~mag and $0.8 \leq$ (F475W - F814W) $\leq 2.4$~mag are ``GCs". We assume the F814W magnitudes of these ``GCs"  follow a Gaussian distribution (i.e., the GCLF) that is right-truncated at $26.3$~mag. The resulting MLE of $\mu_{\text{TO}}$ is a non-sensical $43.7\pm7.75$~mag. The inferred $\mu_{\text{TO}}$ is fainter than the canonical $\mu_{\text{TO}}$ with $99\%$ confidence and the point estimate of $43.7$~mag puts an average ``GC" from this catalog outside the observable Universe! This is strong evidence that there is a significant contaminant population among sources brighter than the canonical $\mu_{\text{TO}}$. Therefore, we consider a two-component parametric mixture model combined with physical information to aid the clustering. Table provides the list of parameters and model components for our parametric mixture model.

 
Specifically, we cluster the color-magnitude data with measurement uncertainties considered for sources with $0.8 \leq$ (F475W - F814W) $\leq 2.4$~mag. Suppose that there are $n$ point sources passing the color cuts with magnitude $M_i$ and color $C_i$, $i\in [n]$. Our two-component mixture model assumes that these sources come from a GCLF component and a contaminant component. We model the magnitude distribution of GCs in these sources using a truncated GCLF with parameter $\theta_{\text{GC}}^M = (\mu_{\text{TO}}, \sigma_{\text{GC}})$, which are respectively the GCLF TO and dispersion. For the color of GCs, we model it by a mixture of two Gaussian distributions that represent the blue and red GCs, each of these components have parameters $\theta_{\text{GC,r}}^C = \left(\mu_{\text{GC,r}}^C, \sigma_{\text{GC,r}}^C\right)$ and $\theta_{\text{GC,b}}^C = \left(\mu_{\text{GC,b}}^C, \sigma_{\text{GC,b}}^C\right)$. For the contaminant component, we model its magnitude distribution by a Gaussian with parameter $\theta_{\text{cont}} = (\mu_{\text{cont}}, \sigma_{\text{cont}})$, and the color distribution by a uniform distribution in $(0.8\text{ mag}, 2.4\text{ mag})$. For simplicity, we assume that the color and magnitude distributions of all components are independent. Denote $\mathbf{\Theta}_{\text{mix}} = \left\{w, w_r, \theta_{\text{GC}}^M, \theta_{\text{GC,r}}^C, \theta_{\text{GC,b}}^C, \theta_{\text{cont}}\right\}$ Our two-component  mixture model is thus:
\[
\pi\left(M_i, C_i \mid \mathbf{\Theta}_{\text{mix}} \right) =& w\pi_{\text{GC}}^M\left(M_i; \theta_{\text{GC}}^M\right)\left\{w_r\pi_{\text{GC,r}}^C\left(C_i; \theta_{\text{GC,r}}^C\right) + (1-w_r)\pi_{\text{GC,b}}^C\left(C_i; \theta_{\text{GC,b}}^C\right)\right\} + \\
&(1-w)\pi_{\text{cont}}^M(M_i; \theta_{\text{cont}})\pi_{\text{cont}}^C(C_i) \\
\overset{\triangle}{=}& w\pi_{\text{GC}}\left(M_i, C_i; w_r, \theta_{\text{GC}}^M, \theta_{\text{GC,r}}^C, \theta_{\text{GC,b}}^C\right) + (1-w)\pi_{\text{cont}}\left(M_i, C_i; \theta_{\text{cont}}\right).
\]
In the above, $w \in [0,1]$ is the proportion of GCs, $w_r \in [0,1]$ is the proportion of GCs that are red. Moreover,
\[
\pi_{\text{GC}}^M\left(M_i; \theta_{\text{GC}}^M\right) = \frac{\phi\left(M_i; \mu_{\text{GC}}^M, \sigma_{\text{GC}}^M\right)f(M_i)}{\int\phi\left(m; \mu_{\text{GC}}^M, \sigma_{\text{GC}}^M\right)f(m)dm},
\]
is the GCLF truncated by the completeness fraction $f(m)$. $\phi\left(M_i; \mu_{\text{GC}}^M, \sigma_{\text{GC}}^M\right)$ is the true GCLF of GCs, and we have set $\mu_{\text{GC}}^M = 26.3$~mag (canonical TO), which is the only parameter we fix. For simplicity, we take $f(m)$ to be the average completeness fraction from J24. $\pi_{\text{GC,r}}^C\left(C_i; \theta_{\text{GC,r}}^C\right)$ and $\pi_{\text{GC,b}}^C\left(C_i; \theta_{\text{GC,b}}^C\right)$ are respectively the Gaussian densities for the color of red and blue GCs. $\pi_{\text{cont}}^M\left(M_i; \theta_{\text{cont}}\right)$ is the Gaussian density of the magnitude distribution of the contaminants, while $\pi_{\text{cont}}^C(C_i)$ is the uniform color distribution of the contaminants. We then obtain the maximum likelihood estimate (MLE) of $\widehat{\mathbf{\Theta}}_{\text{mix}} = \left\{\hat{w}, \hat{w}_r, \hat{\theta}_{\text{GC}}^M,  \hat{\theta}_{\text{GC,r}}^C, \hat{\theta}_{\text{GC,b}}^C, \hat{\theta}_{\text{cont}}\right\}$. The probability that the $i$-th source is a GC is then
\[
p_{i, \text{GC}} = \frac{\hat{w}\pi_{\text{GC}}\left(M_i, C_i; \hat{w}_r, \hat{\theta}_{\text{GC}}^M, \hat{\theta}_{\text{GC,r}}^C, \hat{\theta}_{\text{GC,b}}^C\right)}{\hat{w}\pi_{\text{GC}}\left(M_i, C_i; \hat{\theta}_{\text{GC}}^M, \hat{\theta}_{\text{GC,r}}^C, \hat{\theta}_{\text{GC,b}}^C\right) + (1-\hat{w})\pi_{\text{cont}}\left(M_i, C_i; \hat{\theta}_{\text{cont}}\right)}
\]

\begin{figure*}[t]
    \centering
    \subfigure[Magnitude (F814W) distribution and  mixture model result]{\includegraphics[width = 0.45\linewidth]{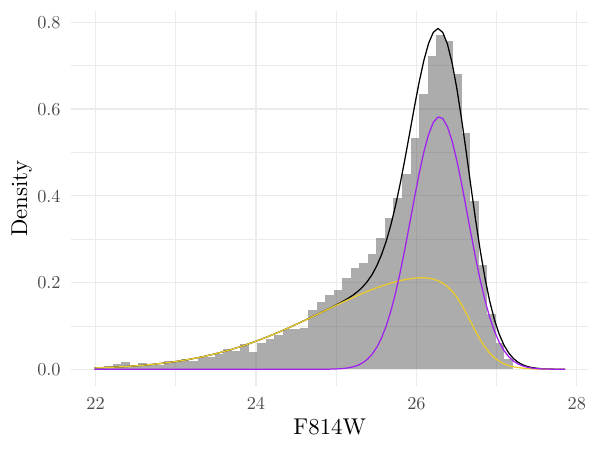}\label{fig:GC_jans_mag_mix}}%
    \subfigure[Color distribution and  mixture model result]{\includegraphics[width = 0.45\linewidth]{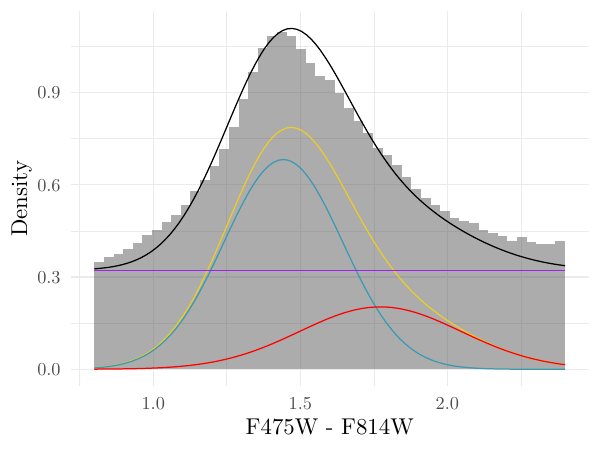}\label{fig:GC_jans_color_mix}}%
    \caption{(a) Magnitude and (b) color distribution (grey histograms) of jittered sources from $500$ simulations using measurement uncertainties and data from J24. A color cut of $0.8 \leq$ F475W - F814W $\leq 2.4$ is applied to the jittered sources. (a) The black line is the best-fit two-component  mixture model for the magnitude distribution of sources. The yellow line is the best-fit truncated GCLF component. The purple line is the best-fit Gaussian distribution for the contaminants component at fainter levels. (b) The black line is the best-fit two-component  mixture model for the color distribution of the sources. The yellow line is the best-fit GC color component. The blue and red lines are GC color distributions of blue and red GCs. The purple line is the color distribution of the contaminant. All best-fit lines are the average results from the simulations.}
    \label{fig:GC_mixture_Jans}
\end{figure*}

We apply the procedure $500$ times to sources that pass the color cuts after jittering the colors and magnitudes of all sources in J24 by their measurement uncertainties.

Figure \ref{fig:GC_mixture_Jans} shows the magnitude and color distributions of the uncertainty jittered sources from the $500$ simulations, as well as the best fit mixture model. In both Figures \ref{fig:GC_jans_mag_mix} and \ref{fig:GC_jans_color_mix}, the black lines are the best-fit overall two-component mixture results. The yellow lines are the truncated GCLF component, while the purple lines are the contaminant component. In Figure \ref{fig:GC_jans_color_mix}, the red and blue lines are the respective red and blue GC sub-populations. The mean color estimates are $\sim 1.44$~mag for blue GCs and $\sim1.77$~mag for red GCs, and they are consistent with the estimates in \cite{Harris2020}. However, roughly half of the sources with $0.8 \leq$ (F475W - F814W) $\leq 2.4$~mag are contaminant, and most of them have F814W magnitudes fainter than $25.5$~mag. Additionally, the contaminant population has the highest density at exactly the canonical GCLF TO point of $26.3$~mag. Therefore, it is nearly impossible to separate GCs from contaminants in this region. This is also demonstrated in the right panel of Figure \ref{fig:GC_prob} where the probabilities of sources being GCs drop significantly in said region.

As a final note, we also examined the magnitude distribution of our data from DOLPHOT in a similar fashion. We found that there is no significant contaminant population in our data, as the magnitude distribution of sources that pass the color cuts almost coincides exactly with the truncated canonical GCLF. This shows that our catalog is rather clean at fainter levels. In this case, the previous non-parametric finite mixture model will work well at distinguishing GCs.

\section{Data Generating Process for Point Sources}\label{sec:DGP pp}

To incorporate the probabilistic GC catalog into our model, additional structure is required to conduct inference. We denote the extracted point sources with point pattern $\mathbf{x} = \{x_1, \dots, x_n\}$, their magnitudes $\mathbf{M} = \{M_1, \dots, M_n\}$, and their colors $\mathbf{C} = \{C_1, \dots, C_n\}$. Let $\mathbf{Z} = (Z_1, \dots, Z_n)$ be a random vector with $Z_i=1$ denoting the $i$-th source is a GC while $Z_i = 0$ being the $i$-th source is not a GC.

Assume that a data-generating process $\mathbb{M}_{\mathrm{GC}}$ with parameters $\boldsymbol{\Theta}_{\mathrm{GC}}$ gives rise to the GC point pattern $\mathbf{x}_{\mathrm{GC}}$ and magnitude $\mathbf{M}_{\mathrm{GC}}$, while another data-generating process $\mathbb{M}_{\mathrm{NG}}$ with parameters $\boldsymbol{\Theta}_{\mathrm{NG}}$ generates point pattern $\mathbf{x}_{\mathrm{NG}}$ and magnitude $\mathbf{M}_{\mathrm{NG}}$ for sources that are not GCs. We assume that $Z_i$'s are conditionally independent given the data and model parameters. The full posterior is then
\begin{equation}\label{eqn:full_post}
\pi(\boldsymbol{\Theta}_{\mathrm{GC}},\boldsymbol{\Theta}_{\mathrm{NG}},\mathbf{Z} \mid \mathbf{x}, \mathbf{M}, \mathbf{C}).
\end{equation}

Naturally, the posterior in Eq.~\ref{eqn:full_post} can be sampled from the following data-augmentation scheme \citep{vandyk_2001, neal_exact_2015}:
    \begin{enumerate}
    \item \label{item: DA step 1}  Sample $\mathbf{Z}' \sim \pi(\mathbf{Z} \mid \mathbf{x}, \mathbf{M}, \mathbf{C}, \boldsymbol{\Theta}_{\mathrm{GC}},\boldsymbol{\Theta}_{\mathrm{NG}})$,
    \item \label{item: DA step 2} Sample $\boldsymbol{\Theta}_{\mathrm{GC}}',\boldsymbol{\Theta}_{\mathrm{NG}}' \sim \pi(\boldsymbol{\Theta}_{\mathrm{GC}},\boldsymbol{\Theta}_{\mathrm{NG}} \mid \mathbf{Z}', \mathbf{x}, \mathbf{M})$.
    \end{enumerate}
Note that the conditional posterior in Step~\ref{item: DA step 2} above does not depend on $\mathbf{C}$ as $(\boldsymbol{\Theta}_{\mathrm{GC}},\boldsymbol{\Theta}_{\mathrm{NG}})$ are unrelated to the color. However, in Step~\ref{item: DA step 1}, sampling $\mathbf{Z}'$ requires parametric modelling of the spatial distribution of non-GC point sources and the entire CMD. Such task is extremely difficult since non-GC sources encompass various objects including foreground stars and background galaxies. These non-GC sources have extremely complex distribution and it is simply impossible to parametrically model both their spatial distributions and CMDs.

One key observation is that if we know the nature of each point source, i.e., $\mathbf{Z}$, the resulting conditional posterior of $\boldsymbol{\Theta}_{\mathrm{GC}}$ and $\boldsymbol{\Theta}_{\mathrm{NG}}$ are independent. Thus,
\[\label{eqn:conditional_independence}
    \pi(\boldsymbol{\Theta}_{\mathrm{GC}},\boldsymbol{\Theta}_{\mathrm{NG}} \mid \mathbf{Z}, \mathbf{x}, \mathbf{M}) =& \pi_{\mathbb{M}_{\mathrm{GC}}}(\boldsymbol{\Theta}_{\mathrm{GC}} \mid \mathbf{Z}, \mathbf{x}, \mathbf{M}) \pi_{\mathbb{M}_{\mathrm{NG}}}(\boldsymbol{\Theta}_{\mathrm{NG}} \mid \mathbf{Z}, \mathbf{x}, \mathbf{M}),
\]
and we can sample the conditional posterior of $\boldsymbol{\Theta}_{\mathrm{GC}}, \boldsymbol{\Theta}_{\mathrm{NG}}$ separately. Indeed, we do not even require the posterior of $\boldsymbol{\Theta}_{\mathrm{NG}}$. However, if we ignore the modeling and sampling of $\boldsymbol{\Theta}_{\mathrm{NG}}$, the conditional distribution for $\mathbf{Z}$ is unavailable. Hence, we consider a Bayesian cut model \citep{plummer_cuts_2015, taborsky2021}, where we introduce additional variables $\mathbf{p}(\mathbf{C}, \mathbf{M}) = \{p(C_i, M_i)\}_{i=1}^n$, on which $\mathbf{Z}$ solely depend, while cutting the dependence of $\mathbf{Z}$ on all other model components:
\[
\pi(\mathbf{Z} \mid \mathbf{x}, \mathbf{M}, \mathbf{C}, \boldsymbol{\Theta}_{\mathrm{GC}},\boldsymbol{\Theta}_{\mathrm{NG}}) \approx \pi(\mathbf{Z} \mid \mathbf{p}(\mathbf{C}, \mathbf{M})).
\]
Under such a cut model, we have
\[
Z_i \sim \mathrm{Bernoulli}(p(C_i, M_i)),
\]
where $p_i = p(C_i, M_i)$ is the probability that the $i$-th source is a GC.  $\mathbf{p}(\mathbf{C}, \mathbf{M})$ is then determined by the finite-mixture models fitted to the CMD shown in Figure \ref{fig:GC_prob}. Thus, inference is facilitated through the following adjusted data-augmentation scheme:
\begin{enumerate}
    \item Sample $\mathbf{Z}' \sim \pi(\mathbf{Z} \mid \mathbf{p}(\mathbf{C}, \mathbf{M}))$,
    \item Sample $\boldsymbol{\Theta}_{\mathrm{GC}}' \sim \pi_{\mathbb{M}_{\mathrm{GC}}}(\boldsymbol{\Theta}_{\mathrm{GC}} \mid \mathbf{Z}, \mathbf{x}, \mathbf{M})$.
\end{enumerate}
Note that the spatial information $\mathbf{x}$ of point sources is not utilized to inform the nature of point source as it may introduce selection bias. For example, in \cite{Carlsten_2022}, they deemed sources close to a galaxy are more likely to be GCs. However, as seen in \cite{Li2022, li2024poisson, vanDokkum_2024} with the discovery of potentially dark galaxies such as CDG-1 and CDG-2, it is entirely possible that GCs can belong to almost dark galaxies that were not discovered yet. Thus, considering the spatial distribution of point sources and their proximity to known galaxies can bias the classification results.

\section{Priors}\label{sec:prior}

We now outline the prior distributions for the model parameters. For the mean count of GCs, $\varlambda_g$, in elliptical galaxies, we assign an informative prior as follows:
$$
\log(\varlambda_g) \sim \mathcal{N}(\log(N_{\mathrm{SF}}^g), 0.25^2).
$$
Here, $N_{\mathrm{SF}}^g$ is determined using specific frequency ($S_N$) relations \citep{Harris1991}:
$$
S_N = N_{\mathrm{GC}}10^{0.4(M_V + 15)}.
$$
$S_N$ represents the specific frequency, and $N_{\mathrm{GC}}$ is the number of GCs in the galaxy. For this study, we set $S_N = 2$, a value expected for typical elliptical galaxies \citep{Harris1991}. The standard deviation of $0.25$ serves as an appropriate representation of the uncertainty observed in GC counts.

For UDGs and LSBGs, we assume a weakly-informative folded Gaussian distribution for $\varlambda_u$:
$$
\varlambda_u \sim \mathrm{f}\mathcal{N}(0, 50^2),
$$
A folded Gaussian distribution is obtained by taking absolute value of a Gaussian random variable. The standard deviation of $50$ signifies the currently observed upper bound of $\sim 100$ GCs in UDGs.

For half-number radii ($R_h^g$) of GC systems in elliptical galaxies, we leverage a relationship obtained from \cite{Forbes2017}. This relationship links the half-number radius of GC systems in elliptical galaxies to their half-light radius, $R_{\mathrm e}^g$. Specifically, it follows that $R_h \sim 3.7R_{\mathrm e}$. $R_{\mathrm e}^g$ of bright elliptical galaxies from our data is obtained through SExtractor \citep{Bertin1996, Li2022}. Therefore, we assign the following log-normal prior to $R_h^g$:
$$
\log(R_h^g) (\log(\mathrm{kpc})) \sim \mathcal{N}(\log(3.7R_{\mathrm e}^g), 0.25^2).
$$
The choice of a standard deviation of 0.25 in the above equation appropriately represents the observed uncertainty associated with this relationship.

For $R_h^u$ of UDG or LSBG GC system, previous studies have found that $R_h^u \sim 0.8R_{\mathrm e}^u - 1.5R_{\mathrm e}^u$ with $R_{\mathrm e}^u$ the half-light radius of a UDG \citep{van_Dokkum_2017, Lim2018, Saifollahi2022}. Thus, we set
$$
\log(R_h^u) (\mathrm{kpc}) \sim \mathcal{N}(\log(R_{\mathrm e}^u), 0.5^2).
$$
The values of $R_{\mathrm e}^u$ for all LSBGs in our data follow from J24. There are a few LSBGs that do not have an estimated half-light radius from J24 due to their small size or faintness. For these ones, we simply set $R_{\mathrm e}^u = 1.5$~kpc.

For the $\mathrm{S\acute{e}rsic}$ index, we prescribe the following prior distributions:
For elliptical galaxies:
$$
\log(\alpha_g) \sim \mathcal{N}(\log(0.5), 0.5^2).
$$
For UDGs:
$$
\log(\alpha_u) \sim \mathcal{N}(\log(1), 0.75^2).
$$
Since all galaxies analyzed in our dataset fall within the lower-mass spectrum, their GC systems generally exhibit smaller $\mathrm{S\acute{e}rsic}$ indices when compared to the most massive giant elliptical galaxies, which can exhibit $\mathrm{S\acute{e}rsic}$ indices exceeding $4$.

For the parameters $\mu_{\text{TO}}^k$ and $\sigma_k$ of the GCLF, we assign the following prior distributions:
$$
\mu_{\text{TO}}^k (\mathrm{mag}) \sim \mathcal{N}(26.3, 0.5^2)
$$
and
$$
\log(\sigma_k) (\mathrm{mag}) \sim \mathcal{N}(\log(1.3), 0.25^2).
$$
The TO point of the GCLF $\mu_{\text{TO}}^k$ is given a Gaussian prior with mean centered at the canonical GCLF TO of $26.3$~mag \citep[][J24]{Harris2020} and standard deviation of $0.5$~mag. The provided standard deviation is intentionally set to a larger value to allow for potential variation of the GCLF TO in UDGs. The GCLF dispersion is given a log-normal distribution with mode at $1.2$~mag, which is also the dispersion of the canonical GCLF \citep[cf.][]{Harris2020}.

Lastly, for the background GC intensity $\lambda_0$, we assign it a log-normal distribution as follows:
$$
\log(\lambda_0) \sim \mathcal{N}(\log(\ell_0), 0.4^2).
$$
The value of $\ell_0$ is image dependent: we determine $\ell_0$ by first subtracting the number of GCs in galaxies (assumed as the prior median of $\varlambda_k$) from the total number of GCs in an image, ensuring it roughly coincides with physical reality. We then obtain a final value of $\ell_0$ by accounting for unobserved GCs using the canonical GCLF. The standard deviation of $0.4$ is selected to accommodate the substantial uncertainty associated with this process.

\section{Computation and Inference}\label{sec:inference}

To conduct inference for our model, we employ an adaptive MCMC \citep{haario2001, roberts2009} algorithm in conjunction with a data-augmentation scheme \citep{vandyk_2001, backlund_overview_nodate}.  

Following Section \ref{sec:DGP pp}, we incorporate a data-augmentation scheme to facilitate inference with $\mathbf{Z}$ being the auxiliary variable. We then generate $\boldsymbol{\Theta}$ based on $\pi(\boldsymbol{\Theta} \mid \mathbf{Z}, \mathbf{x}, \mathbf{M})$ using an adaptive MCMC algorithm.

For the adaptive MCMC transition kernel $p(\cdot \mid \boldsymbol{\Theta}_i)$, we consider the following
\begin{equation}\label{eqn:propsal}
  \boldsymbol{\Theta}_{i+1}\mid \boldsymbol{\Theta}_i \sim
    \begin{cases}
      \mathcal{N}(\boldsymbol{\Theta}_i, C_\theta) & \text{$i \leq 1000$},\\
      \mathcal{N}(\boldsymbol{\Theta}_i, \gamma\mathrm{Cov}(\{\boldsymbol{\Theta}_j\}_{j=1}^i)) & \text{$i > 1000$},
    \end{cases}       
\end{equation}
$C_\theta$ here is a diagonal matrix and $\gamma > 0$ is a scaling parameter. For different images, $C_\theta$ is different and fine-tuned. Following \citep{haario2001, roberts2009}, the optimal value of $\gamma$ is set to $\gamma = 2.38^2/\text{dim}(\mathbf{\Theta})$. Note that the parameters in Eq.~\ref{eqn:propsal} are transformed so the proposed values would not cause issues in computing the model likelihood.

\bibliography{bibliography}{}
\bibliographystyle{aasjournal}


\end{CJK}
\end{document}